\documentclass[prd, amsfonts, twocolumn, nofootinbib, showpacs]{revtex4-1}
\usepackage{graphicx, epsfig}
\usepackage{color}
\usepackage{amsmath}
\usepackage{mathtools}
\newcommand{\be}{\begin{equation}}
\newcommand{\ee}{\end{equation}}
\newcommand{\bea}{\begin{eqnarray}}
\newcommand{\eea}{\end{eqnarray}}

\newcommand{\gapp}{\mathrel{\raise.3ex\hbox{$>$}\mkern-14mu \lower0.6ex\hbox{$\sim$}}}
\newcommand{\lapp}{\mathrel{\raise.3ex\hbox{$<$}\mkern-14mu \lower0.6ex\hbox{$\sim$}}}
\newcommand{\LSM}{L$\Sigma$M}
\newcommand{\LSMmth}{GML}
\newcommand{\GMLfull}{Gell Mann-L\'evy~}
\newcommand{\GML}{GML~}

\newcommand{\HVEV}{\langle H\rangle}
\newcommand{\Tree}{Tree}
\newcommand{\half}{\frac{1}{2}}
\newcommand{\mpisq}{m_\pi^2}
\def\bbox{{\,\lower0.9pt\vbox{\hrule \hbox{\vrule height 0.2 cm
\hskip 0.2 cm \vrule  height 0.2 cm}\hrule}\,}}

\begin{document}
\title{The ``Goldstone Exception" II: 
 Absence of a Higgs Fine-Tuning Problem in the 
\\ Spontaneously Broken Limit  of the \GMLfull Linear Sigma Model:
\\ O(4) with Partially Conserved Axial-vector Currents (PCAC) and
\\  ${\bf {\rm SU(2)_L}}$ with PCAC and Standard Model Quarks and Leptons}
\author{Bryan W. Lynn$^{1}$, Glenn D. Starkman$^{2,3}$, Katherine Freese$^4$ and Dmitry I. Podolsky$^2$  }
\affiliation{$^1$ University College London, London WC1E 6BT, UK}
\affiliation{$^2$ ISO/CERCA/Department of Physics, Case Western Reserve University, Cleveland, OH 44106-7079}
\affiliation{$^3$ Theory Group, Physics Department, CERN, CH-1211 Geneva 23, Switzerland}
\affiliation{$^4$ Department of Physics, University of Michigan, Ann Arbor, MI }
\email{bryan.lynn@cern.ch, glenn.starkman@case.edu,\newline ktfreese@umich.edu, podolsky@phys.cwru.edu}
%

\begin{abstract}

More than four decades ago, B.W.~Lee and K.~Symanzik proved (but did not say) that, in the \GMLfull model with conserved vector currents (CVC) and partially conserved axial-vector currents (PCAC) (i.e. a generic set of  global O(4) linear sigma models (\LSM) in the $\vert\HVEV\vert$ vs. $\mpisq/\lambda^2$   quarter-plane), tadpole renormalization (a Higgs Vacuum Stability Condition) 
forces all S-matrix ultra-violet quadratic divergences (UVQD)  to be absorbed into the physical renormalized pseudo-scalar pion (pole) mass squared, $\mpisq$. We show that this includes ``new" UVQD (widely unfamiliar to modern audiences) which we identify as corrections to the PCAC relation $\partial_\mu {\vec J}^5_\mu = -\HVEV \mpisq \vec \pi$. We also show that tadpole renormalization is an automatic consequence of Ward-Takahashi identities. 

{\em We prove that all UVQD therefore vanish identically in the ÒGoldstone-modeÓ  limit, $\mpisq\to m^2_{\pi;NGB;\LSMmth}\equiv0$,
where pions are Nambu-Goldstone Bosons (NGB),}
and where Lee and Symanzik's Goldstone Symmetry Restoration Condition (a renormalization prescription) enforces spontaneous symmetry breaking and the {\em exact} massless-ness of NGB as required by Goldstone's theorem.
Axial-vector current conservation is restored $\partial_\mu {\vec J}^5_\mu\to0$ in the spontaneously broken limit  of the \GMLfull \LSM~ and, by consequence, so is   $SU(2)_{L-R}$ chiral  symmetry: {\em the vanishing of UVQD is therefore achieved in the Goldstone-mode by restoration of an exact symmetry, and therefore (by definition) without fine-tuning!}
A weak-scale Higgs mass is therefore not UVQD fine-tuned in the spontaneously broken \GML \LSM.  That fact is simply another (albeit unfamiliar) consequence of the Goldstone Theorem, an exact property of the spontaneously broken  vacuum and spectrum.
Hence Goldstone-mode O(4) \LSM~ symmetries are sufficient to ensure that the theory 
does not suffer from the Higgs Fine Tuning Problem or Naturalness Problem.   
This is contrary to the widely accepted belief that UVQD in the Higgs mass, arising already at 1-loop, lead to such problems in the O(4) \LSM,
which are then presumed to be inherited by the Standard Model (SM). 
The key observation is to  regard the spontaneously broken O(4) \LSM~ as the Goldstone-mode limit of the \GMLfull \LSM~ with PCAC.

We prove this first at 1-loop for the pure scalar \GMLfull model, then extend the proof to all-orders of loop perturbation theory. We then break the symmetry $O(4) \to SU(2)_L$ with SM Yukawa couplings,  and show that the above remains true when the \LSM~with PCAC is extended  to include SM quarks and leptons, first to 1-loop then to all-loop-orders.

It was recently shown that, including all-orders perturbative electro-weak and QCD loops, and independent of 1-loop ultraviolet regularization scheme, 
all finite remnants of  UVQD in the Standard Model S-Matrix 
are absorbed into NGB masses-squared  $m_{\pi;NGB;SM}^2$, 
 and vanish identically when $m_{\pi;NGB;SM}^2\to0$ , i.e. in the (spontaneously broken) SM. 
The SM Higgs mass is therefore not fine-tuned.  This paper's results on the spontaneously broken limit of the {\em un-gauged} $SU(2)_L$ \LSM~with SM fermions  are the naive-zero-gauge-coupling limit (and constitute a profound simplification)  of that SM result, which arises from the embedding of the $\it gauged$ $SU(2)_L$ \LSM~ into the spontaneously broken SM.
\end{abstract}

\maketitle
\section{Introduction}

The idea that ultra-violet quadratic divergences (UVQD) in the Standard Model (SM) require extreme fine-tunings
in order to result in weak scale physics, 
has had enormous impact in motivating the possible existence of Beyond the  Standard Model physics,
such as supersymmetry, technicolor,  large extra dimensions, {\it etc}.   These UVQD are canonically
demonstrated  (at one-loop with UV cutoff $\Lambda$) in the O(4) \LSM~ ({\it eg.} \cite{Susskind1979}), which is embedded in the electroweak sector of the SM.
It is then argued that these ${\cal O}(\Lambda^2)$ contributions to, say the Higgs mass-squared, 
must be  fine-tuned    against similarly large counter-terms to obtain a small ${\cal O}(10^4 GeV^2)$ 
Higgs mass-squared (within the weak scale \LSM) and similar magnitude weak gauge boson masses. 

In this paper, we will show that, {\em contrary to this lore}, the \GMLfull (\GML) linear sigma model \cite{GellMannLevy1960} of scalars with Partially Conserved Axial-vector Currents (PCAC) \cite{AdlerDashen1968} is perfectly
well behaved in its spontaneously broken limit.
No fine-tuning is necessary to keep the Higgs mass small in \GML 
``Goldstone mode,'' where the O(4) symmetry is spontaneously broken (and axial-vector currents are conserved), leaving behind three precisely massless 
Nambu-Goldstone Bosons (NGB)\cite{Nambu1959,Goldstone1961,Goldstone1962}. This is because all UVQD are contained in the mass squared of the NGB,
so enforcing the Goldstone Theorem (i.e. insisting that the NGBs are indeed massless) \cite{Goldstone1961,Goldstone1962} eliminates all UVQD, including 
that in the Higgs mass, without fine-tuning.  

The UVQD of the  \GMLfull Goldstone mode (with reduced symmetry $O(4) \to SU(2)_L$ due to SM Yukawa couplings)  are precisely the naive zero-gauge-coupling limit of the UVQD in the {\em gauged} $SU(2)_L$ \LSM~ embedded in the SM, 
where those three massless Nambu Goldstone bosons are ``eaten" (within the Higgs mechanism) by the three gauge bosons of the three broken generators of the $SU(2)_L \times U(1$) symmetry.

We show that the O(4) \LSM~ theory of scalars does not suffer from a Higgs-FTP in Goldstone mode. We then extend this result to include Standard Model quarks and leptons.

On a separate issue, the Higgs Vacuum Stability Condition (Higgs-VSC) ensures that the physical vacuum
is stable against the spontaneous creation or disappearance of physical Higgs particles.
Lee \cite{Lee1970} enforces that stability condition (by hand) throughout the $\vert\HVEV\vert$ vs. $\mpisq / \lambda^2$ quarter-plan, including the Goldstone mode. We show instead that the Higgs-VSC is an automatic consequence of Ward-Takahashi identities.

{\bf The goal of this paper is to provide details of the UVQD renormalization of the global O(4)\LSM.} We provide a simple and brief illustration of  our basic approach
to the Higgs-FTP problem.
{\bf A subsequent paper \cite{LynnStarkman2013a} gives a more complete yet intuitive explanation, but within the context of the simpler global U(1) \LSM~with a partially conserved axial-vector current (PCAC)}: e.g. including fermions in the UVQD-corrected Ward-Takahashi identities used to  set the strength of the PCAC relation; using Ward-Takahashi identities to show the automatic enforcement of the Higgs-VSC in global theories. It also carefully defines B.W. Lee and K. Symanzik's Goldstone Mode Renormalization Prescription (GMRP, our name) for theories with global symmetries.
 As necessary, we extend here various U(1) results (from \cite{LynnStarkman2013a}) to the case of O(4) studied in this paper: e.g. we prove here the automatic enforcement of the O(4) Higgs-VSC by the Ward-Takahashi identities of B.W. Lee and K. Symanzik \cite{Lee1970,Symanzik1970a,Symanzik1970b,Vassiliev1970}, and display the ``new" UVQD corrections to the O(4) PCAC relation.

In \cite{LynnStarkman2013a}, we also identify a ``Higgs Fine Tuning Discontinuity" (Higgs-FTD): i.e. that the UVQD structure of the Schwinger ($\epsilon \equiv 0$) linear sigma model  \cite{Schwinger1957} is {\em not the same} as the spontaneously broken limit of the UVQD structure of the  \GMLfull ($\epsilon \neq 0$) \LSM. For example, ``new" UVQD corrections (widely unfamiliar to modern audiences and identified as corrections to the PCAC relation) arise in the \GML \LSM, but not in the Schwinger \LSM. The Higgs-FTD is reminiscent of other such  discontinuities: theories of  explicit photon masses \cite{vanDamVeltman1970} and of explicit graviton masses \cite{vanDamVeltman1970,Zakharov1970}  (the so-called vDVZ discontinuity) in which the correct number of degrees of freedom to reproduce the exactly massless theory is not recovered in the massless limit, with phenomenological consequences; or in non-Abelian gauge theories of massive spin-1 bosons, in which explicit mass terms famously generate non-renormalizable  UVQD structure and non-unitary high energy behavior \cite{JCTaylor1976}, while the Higgs mechanism does not. 

The GMRP resolves the Higgs-FTD. 
The GMRP is further extended in Ref. \cite{LynnStarkman2013a} to the UVQD renormalization of the BRST-invariant spontaneously broken U(1) ``Abelian Higgs" gauge theory, resolving its Higgs-FTD and, at the same time, laying a solid foundation for the GMRP-based renormalization of the Standard Model, as discussed and advocated by J.C. Taylor \cite{JCTaylor1976}. The SM GMRP  largely explains the results of \cite{Lynn2011}, which  resolves the Higgs-FTD lurking inside the SM, and thus proves the absence of a Higgs Fine Tuning Problem or Naturalness Problem in the Standard Model.

As a result of the underlying O(4) invariance of the theory,
 the Higgs and pions masses are related by
\begin{equation}
\label{basic}
m_h^2 = \mpisq + 2 \lambda^2 \HVEV^2 
\end{equation}
where we have expanded the Higgs about its vacuum $H = h + \HVEV$.
Note that throughout this paper, we will follow the language of early literature on this subject 
and take the coupling constant for the quartic coupling to be $\lambda^2$ rather than the modern \cite{Ramond2004} convention $\lambda$.

As shown by B.W. Lee, K. Symanzik and A.Vasiliev \cite{Lee1970,Symanzik1970b,Vassiliev1970,Symanzik1970a} and the textbook by C. Itzykson and J-B. Zuber (\cite{ItzyksonZuber1980} equation 11-176),  the value of $\HVEV$ is renormalized only logarithmically (not by UVQDs).
Hence from (\ref{basic}) we see that the divergences of $m_h^2$ are identical to those of $m_\pi^2$. 
Now we take the Goldstone limit of $m_\pi^2 \rightarrow 0$.   Because a new $SU(2)_{L-R}$  chiral symmetry appears at $m_\pi^2=0$,
this limit  is {\em not a fine-tuning}.
 Equation (\ref{basic}) now becomes $m_h^2 = 2 \lambda^2 \HVEV^2 $, which has  no UVQDs.  
 Thus Goldstone's theorem together with the Ward-Takahashi identities have brought about the
 proper value of the Higgs mass.   
  
  Our result depends crucially on using the correct version of the theory.  
  As also shown by Lee,  Symanzik  and  Itzykson and Zuber, the Goldstone mode must properly be
 obtained by starting from a theory with 
  an explicit symmetry breaking term in the Lagrangian $L^{ PCAC}_{\LSMmth} = \epsilon H$.  Then the Ward-Takahashi Identities for the partially
 conserved vector current  $\partial_\mu \vec{J}_\mu^5 = -\HVEV m_\pi^2 \vec \pi$ demand that $\epsilon = m_\pi^2 \HVEV$.
 Thus, not surprisingly, the symmetry breaking term both depends on and is the source of  the explicit pion mass.
 As further shown by Lee,  Symanzik and Itzykson and Zuber, only by considering $\epsilon \neq 0 $, may one start in the Wigner Mode
 ($\HVEV=0$, $\mpisq\neq0$), where renormalizability is proven, and then take a path, {\em analytic in $\mu^2$}, to
 the Goldstone limit with $\epsilon =0$ and $m_\pi^2 = 0$, but $\HVEV\neq0$. This forces the vacuum to be stationary at $H=\HVEV$ and $\langle h \rangle =0$ along that entire analytic path.
 In the Goldstone mode
 the Higgs mass  then naturally takes its correct value without any fine-tuning of  UVQD.

Although the recent trend has been to argue that the SM is only a low-energy effective theory of some more complete theory,
our goal is to understand the properties of the actual SM.   If the SM lacks fine-tuning problems related to ultra-violet quadratic divergences,
then clearly the onus is on those proposing any  extension of the SM to demonstrate that they maintain that Higgs-FTP-free existence; 
for it would not be a problem of the SM, but of the extension.   Because we are interested in the disposition of the
UVQD of the SM, we turn our attention in this paper to the much simpler O(4) \LSM~ and then to the $SU(2)_L$ \LSM~ with SM quarks and leptons: the latter having the same UVQD structure as the SM in the appropriate zero gauge coupling limit.
The taming of logarithmic divergences and the calculation of finite renormalization contributions is not the subject of this paper, 
and it is important that those {\em are} sensitive to the difference between the SM and the O(4) \LSM.
Because we are interested in the disposition only of UVQD  arising from quantum loops in the O(4) \LSM,
and not in the logarithmic divergences or finite parts,
we are able to treat the O(4) \LSM~ in isolation, as a {\bf stand-alone} flat-space quantum field theory:
\begin{itemize}
\item It is not embedded or integrated into any higher-scale ``Beyond the O(4) \LSM''  physics.
\item Loop integrals are cut off at a short-distance UV scale, $\Lambda$.
\item Although the cut-off can be taken to be near the Planck scale $\Lambda\simeq M_{Pl}$ , quantum gravitational loops are not  included.
\end{itemize}

For pedagogical simplicity, we will ignore in all calculations, discussions and formulae  any logarithmic divergences and (except for the pion mass) finite contributions from loops,
as these are unnecessary to (and distracting from) our explanation of the absence of any fine-tuning demanded by UVQD.
Thus, while  for coefficients of dimension-2 relevant operators we will carefully distinguish between renormalized values of quantities such as  $\mu^2$
and the corresponding UVQD bare term $\mu_{Bare}^2$,
we will {\em not} distinguish between things that are only logarithmically or finitely renormalized, 
such as between bare fields and dimensionless couplings and their renormalized values.
For the coefficient of the dimension-1 relevant symmetry breaking PCAC operator, we must therefore distinguish between the quadratically divergent $\epsilon^{Bare}_{\LSMmth}$ and the renormalized $\epsilon^{1-loop}_{\LSMmth}$.
We will also drop all vacuum energy and vacuum bubble contributions as beyond the scope of this paper. 

This paper concerns stability and symmetry restoration protection against only UVQD.
It does not address any of the other, more usual, stability issues  of the SM (cf. the discussion in, for example, \cite{Ramond2004}, and references therein)
or the O(4) \LSM~ (in part because the solutions to those issues are generically different in the SM
than in the O(4) \LSM).

In a recent paper \cite{Lynn2011}, it was shown that, including all-orders perturbative electro-weak and QCD loops, and independent of 1-loop UV regularization scheme, all finite remnants of ultra-violet quadratic divergences (UVQD) in the Standard Model (SM) S-Matrix are absorbed into Nambu Goldstone Boson (NGB) mass-squareds $m_{\pi;NGB;SM}^2$   and vanish identically when $m_{\pi;NGB;SM}^2\to0$ , i.e. in the spontaneously broken SM. 
Therefore, the SM Higgs mass is not fine-tuned. 
But, as a hugely successful theory of Nature, the SM is necessarily quite baroque, and sometimes even opaque. 
In this paper, we derive results in the simpler \GMLfull linear sigma model, ignoring most SM parameters, gauge bosons and ghosts and 
taking the zero-gauge-coupling limit in 1PI 1-loop and multi-loop nested UVQD. 
 This paper's results on the {\it un-gauged} spontaneously broken $SU(2)_L$ \LSM~ are the naive zero-gauge-coupling limit (and constitute a profound simplification)  of the SM results in \cite{Lynn2011}, 
 which arise from the embedding of $\it gauged$ $SU(2)_L$ \LSM~ into the spontaneously broken SM.

Fortunately  for most readers, all relevant O(4) and $SU(2)_L$  \LSM~ (and SM) 1-loop Feynman diagrams were calculated and agreed on long ago.
We refer the interested reader to that vast literature 
\cite{Veltman1981,Passarino1979,Consoli1979,LEPYellow,Polarization,LynnStuart1985,Ellis1996,Ellis1996b,Marciano2000,Levinthal1992,Kennedy1989,TestsofEW,LynnPeskinStuart,LEPStudy,Lynnunpub,Hollik1986,Marciano1975,Marciano1981,Marciano1980,Sirlin1980,Grzadkowski1987,Wetzel1986,Lynn1982,Marciano1982,Lynn1993,Bardin1982,Bardin1980,Bohm1986,Sirlin1986,KennedyLynn1988,Lynn1982b,Stuart1985, Bohm1991,Kinoshita1959,Lee1970,Symanzik1970a,Symanzik1970b,Vassiliev1970,Symanzik1969,Gervais1969,Lee1969,DeGrassiSirlin1992,KurodaMoultakaSchildknecht1991,tHooft1971a,tHooft1971b,Lee1972,LeeZinnJustin1972,LeeZinnJustin1973,tHooftVeltman1972a,tHooftVeltman1972b,ZinnJustin1975,KlubergSternZuber1975a,KlubergSternZuber1975b,Feynman1963,DeWitt1967a,DeWitt1967b,FadeevPopov1967,Mandelstam1967a,Mandelstam1967b,FradkinTyutin1970,Booulware1970,JCTaylor1971,Slavnov1972,SalamStrathdee1970,EnglertBroutThiry1966,BecchiRouetStora1976,Tyutin1975,IofaTyutin1976,Siegel1979,
CapperJonesVanNieuwenhuizen1980,PeskinLynn1984,Lynn1988,Kennedy1988,EinhornJonesVeltman1981,PeskinTakeuchi1990a,KennedyLangacker1990,PeskinTakeuchi1990b,GoldenRandall1990,HoldenTerning1990,AltarelliBarbieri1991,PeskinTakeuchi1992,AltarelliBarbieri1992,Hollik1996,HighPrecisionEWPhysics2006} 
for specifics. 
Motivated by that literature, we use  the Feynman-diagram naming convention of Ref. \cite{Veltman1981} and Euclidean metric.

The structure of the remainder of this paper is:
\begin{itemize}
\item Section 2 clarifies the correct renormalization of spontaneously broken O(4) \LSM, and the $SU(2)_L$ \LSM~   coupled to Standard Model (SM) quarks and leptons. 
Most of the calculations (if not the effective Lagrangian presentation and some of the specific results) in Section 2A through 2E (and probably 2F) are not new 
and have been common knowledge for more than four decades.   Specifically:
	\begin{itemize}
	\item Section 2A introduces the  bare Lagrangian for the (fermion-free) O(4) \LSM~   in the $\vert\HVEV\vert$ vs. $\mpisq/\lambda^2$  quarter-plane (as required by Lee, Symanzik, Itzyson and Zuber  \cite{Lee1970, Symanzik1970b,Vassiliev1970, ItzyksonZuber1980} for proper 		
	renormalization of the O(4) \LSM), and analyzes it at tree-level.   It shows how Ward-Takahashi identities relate the coefficient of the required explicit breaking term to the NGB mass $\mpisq$
	and the Higgs Vacuum Expectation Value $\HVEV$
	in such a way as to impose the vanishing of tadpoles (and thus vacuum stability).	
	\item Section 2B calculates 1-loop UVQD in 2-point self-energies and 1-point tadpoles and shows that Ward-Takahashi identities 
	force all of these to be absorbed into the physical renormalized pseudo-scalar pion (pole) mass-squared $\mpisq$.  
	The continued vanishing of the tadpoles due to Ward-Takahashi identities means that  the stability of the Higgs vacuum is automatically enforced, 
	and does not require the by-hand imposition of a Higgs Vacuum Stability Condition.
	\item Section 2C defines the Higgs Fine Tuning or Naturalness Problem (Higgs-FTP) which emerges in 1-loop-corrected $\mpisq\neq0$  O(4) \LSM~   and,
	in particular, the $\mpisq\neq0$, $\HVEV\to0$  ``Wigner-mode'' limit. 
	\item Section 2D studies the opposite Goldstone-mode  $\mpisq=0$ $\HVEV\neq0$ limit,  
	and shows how the Lee/Symanzik Goldstone Symmetry Restoration Condition (Goldstone-SRC) embedded there causes all remnants of 1-loop UVQD to cancel and vanish identically as 
	$\mpisq\to m_{\pi;NGB;\LSMmth}^2\equiv0$  (exactly zero!),  and the theory to have no Higgs-FTP there. 
	\item Section 2E extends our bosonic Goldstone-mode O(4) \LSM~  results to all-orders in loop-perturbation theory. 
	\item Section 2F breaks the symmetry $O(4) \to SU(2)_L$ in order to further extend our results to include virtual SM quarks and leptons. 
	\end{itemize}
\item Section 3 presents some conclusions drawn from the arguments and calculations of the other sections.
\item Appendix A shows detailed calculations of 1-loop UVQD in O(4) \LSM~ with virtual bosons. 
\item Appendix B shows details for the extension of our results to all-loop-orders for Goldstone-mode scalar O(4) \LSM.
\item Appendix C shows detailed calculations of 1-loop UVQD in $SU(2)_L$ \LSM~ with virtual SM quarks and leptons. 
\item Appendix D gives M.J.G. Veltman's  relation between 1-loop UVQD in dimensional regularization and UV cut-off regularization \cite{Veltman1981}, and shows that  the vanishing of UVQD in Goldstone-mode of the \GMLfull \LSM~ does not  depend on whether dimensional regularization or UV cut-off regularization is used. 

\end{itemize}

\section{\GMLfull Linear sigma model, with either massive pions $\mpisq\neq0$, or Nambu-Goldstone Bosons where $\mpisq\to{}m_{\pi;NGB;\LSMmth}^2\equiv0$ }
\label{sec:O4LSM}

\subsection{O(4) \LSM~   with PCAC in the $\vert\HVEV\vert$ vs. $\mpisq/\lambda^2$  quarter-plane at tree level}
\label{sbscn:O4LSMquarter-plane}

We follow closely the pedagogy (and much of the notation) of Benjamin W. Lee \cite{Lee1970}
and use the manifestly renormalizable linear representation of  $\Phi$  to make clear
the exact not-fine-tuned cancellation of UVQD. The desire to make direct contact with \cite{Lee1970} and the work on which it was built also leads
us to adopt other notational conventions, such as the choice of $\lambda^2$  in (\ref{eqn:LBare-LSM}), 
rather than the more modern $\lambda$.    We begin with the pure scalar bare Lagrangian:
\begin{eqnarray}
\label{eqn:LBare-LSM}
L^{Bare}_{\LSMmth}&=& -\vert\partial_\mu\Phi\vert^2 - V^{Bare}_{\LSMmth}
\\ V^{Bare}_{\LSMmth} &=& \lambda^2\left(\Phi^\dagger\Phi\right)^2 + \mu_{Bare}^2 \Phi^\dagger\Phi -L^{Bare; PCAC}_{\LSMmth} \nonumber
 \end{eqnarray}
 with \GMLfull's explicit symmetry breaking (pion-mass-enforcing) PCAC term
 \be
 \label{eqn:LPCAC}
 L^{Bare; PCAC}_{\LSMmth}=  \epsilon^{Bare}_{\LSMmth} H
 \ee
 and $\lambda^2 > 0$. 
 
 We want a manifestly renormalizable linear representation for the scalars in order to control UVQD. Textbooks by J.C. Taylor \cite{JCTaylor1976} and H. Georgi \cite{Georgi2009}  review the $SU(2)_L \times SU(2)_R$ symmetry, current algebra, CVC and PCAC structure, boson currents and left-handed quark and lepton currents of the \GMLfull model. Since this is locally isomorphic to O(4), we might choose   
 \begin{eqnarray}
 \Phi_{SU(2)_L \times SU(2)_R} = \frac{1}{\sqrt 2} \left[ H+i{\vec \sigma} \cdot {\vec \pi} \right]
 \end{eqnarray}
with Pauli matrices  $\vec \sigma$.  
But, anticipating symmetry breaking $O(4) \to SU(2)_L$ in subsection \ref{sbscn:withSMQL-noHiggs-FTP} (by SM Yukawa couplings, which give mass splitting within left-handed doublets), it is convenient to choose a complex $SU(2)_L$  Higgs doublet representation for the scalars
\begin{eqnarray}
\Phi &\equiv&  \Phi_{SU(2)_L \times SU(2)_R} \left[\begin{array}{c} 1 \\0\end{array}\right]
\\ &=& \frac{1}{\sqrt{2}} \left[ \begin{array}{c}H+i\pi_3\\ -\pi_2 + i\pi_1\end{array}\right] 
\\ \quad \pi_\pm &\equiv& \ \frac{1}{\sqrt{2}} (\pi_1 \mp i \pi_2)
\end{eqnarray}
which is mapped to an O(4) quartet of real scalars:
\be
\label{eqn:O4Fields}
\Phi^\dagger\Phi = \half (H^2 + \vec{\pi}^2)
\ee
Vacuum energy/bubble contributions will be ignored throughout. 
We assume  $\lambda^2={\cal O}(1)$.
The Lagrangian (\ref{eqn:LBare-LSM}) has three free parameters:  $\lambda^2$, $\mu_{Bare}^2$ and $\epsilon^{Bare}_{\LSMmth}$.

Apart from  the explicit symmetry breaking term  (\ref{eqn:LPCAC}), 
the theory described by (\ref{eqn:LBare-LSM}) has $SU(2)_L \times SU(2)_R \sim O(4)$ symmetry. The $3+3=6$ generators yield vector currents $\vec{J}_\mu$  
and axial-vector currents $\vec{J}_\mu^5$:
\begin{eqnarray}
\vec{J}_\mu &\equiv& \vec{\pi} \times \partial_\mu \vec{\pi} \\
\vec{J}_\mu^5 &\equiv& \vec{\pi}\partial_\mu H - H \partial_\mu\vec{\pi} .\nonumber
\end{eqnarray}
Although the vector current  is conserved (CVC hypothesis), with the explicit symmetry breaking term (\ref{eqn:LPCAC}) the divergence of the axial-vector current is non-zero, so it has only partially conserved axial-vector currents (PCAC hypothesis). In {\em  Euclidean metric}:
\begin{eqnarray}
\label{eqn:AxialCurrentDivergence}
\partial_\mu\vec{J}_\mu^5 &=& -\epsilon^{Bare}_{\LSMmth} {\vec \pi}
\\ \partial_\mu\vec{J}_\mu &=& 0 
\end{eqnarray}

As further shown by B.W. Lee \cite{Lee1970}, K. Symanzik \cite{Symanzik1970a,Symanzik1970b} and C. Itzykson and J-B. Zuber \cite{ItzyksonZuber1980},  the value of $\epsilon^{Bare}_{\LSMmth}$ in (\ref{eqn:LBare-LSM}) is 
determined by the  first of the connected Green's function Ward-Takahashi identities, 
connecting the vacuum with the on-shell one-pion state of momentum $q_\mu$.  This reads 
\cite{Lee1970,Symanzik1970a,Symanzik1970b,Vassiliev1970,
Symanzik1969,Gervais1969,Lee1969,Bogoliubov1957,Hepp1966,ZimmermanDeser,ItzyksonZuber1980}
\be
\label{eqn:WTidentity}
\langle 0 \vert J_\mu^{5,i}(x) \vert \pi^j(q)\rangle \equiv -i\delta^{ij} \HVEV q_\mu {\hat \pi}^i e^{iq_\mu x_\mu} \,,
\ee
where  $\HVEV\equiv\langle0\vert H\vert0\rangle$ is the (renormalized) Higgs  VEV.
The divergence of (\ref{eqn:WTidentity}) is
\begin{eqnarray}
\label{eqn:WTdivergence}
\partial_\mu\langle 0 \vert J_\mu^{5,i}(x) \vert \pi^i(q)\rangle &\equiv& \delta^{ij} \HVEV q^2 {\hat \pi}^i e^{iq_\mu x_\mu}  \\
&=& -\delta^{ij} \HVEV \mpisq {\hat \pi}^i e^{iq_\mu x_\mu}.\nonumber
\end{eqnarray}
The final equality being valid only for an on-shell pion,
$\mpisq\geq0$ is  the physical renormalized pseudo-scalar pion (pole) mass-squared.
Thus, the PCAC relation
\be
\label{eqn:DivergenceofAmu}
\partial_\mu\vec{J}_\mu^5 = -\HVEV\mpisq {\vec \pi}\,.
\ee
Equating coefficients of ${\vec \pi}$ on the right-hand sides of (\ref{eqn:AxialCurrentDivergence}) and (\ref{eqn:DivergenceofAmu}),
we see that $\HVEV$ and the coefficient $\epsilon^{Bare}_{\LSMmth}$ of the symmetry breaking term must be related by
\be
\label{eqn:epsilonequality}
\epsilon^{Bare}_{\LSMmth} = \HVEV \mpisq.
\ee
We can therefore rewrite the bare potential:
\be
V^{Bare}_{\LSMmth} = \lambda^2\left(\Phi^\dagger\Phi\right)^2 + \mu_{Bare}^2 \Phi^\dagger\Phi - \HVEV \mpisq H \,.
\ee

It is straightforward to show that the minimum of $V_{Bare}$ is at 
\begin{equation}
\langle\Phi\rangle = \frac{1}{\sqrt{2}}\left[\begin{array}{c}\HVEV\\0\end{array}\right]\,,
\end{equation}
with the tree-level result
\be
\label{eqn:HVEVsq}
\mu_{Bare}^2 = \mu^2 \equiv \mpisq-\lambda^2 \HVEV^2
\ee

It is then particularly instructive to eliminate the unphysical parameter $\mu^{2}_{Bare}$ from $V^{Bare}_{\LSMmth}$, while holding the physical renormalized pion mass-squared $\mpisq$ fixed, 
\be
\label{eq:form}
V^{Bare}_{\LSMmth} \!=\! \lambda^2\left[ \Phi^\dagger\Phi \!-\! \frac{1}{2}\left(\HVEV^2 \!-\! \frac{\mpisq}{\lambda^2}\right) \right]^2 \!-\! \HVEV \mpisq H \,.
\ee
This is the famous \GMLfull potential \cite{GellMannLevy1960,AdlerDashen1968,Lee1970,Symanzik1970a,Symanzik1970b,Vassiliev1970,JCTaylor1976,ItzyksonZuber1980}. A cross section is plotted in Figure \ref{fig5} where (for this tree-level subsection)  it is called $V^{Tree}_{\LSMmth}$. From this we see very clearly the roles played by the symmetry breaking term $\HVEV \mpisq H$ and by the pion mass-squared in determining
the shape and symmetry of the potential. We have also traded the three free parameters in (\ref{eqn:LBare-LSM}) (physical $\lambda^2$ and un-physical $\mu_{Bare}^2$ and $\epsilon^{Bare}_{\LSMmth}$) for three physical inputs ($\lambda^2$, $\HVEV$ and $\mpisq$).

 We will show below that the form of the \GMLfull potential in (\ref{eq:form}) will remain unchanged, including UVQD to all loop-orders of perturbation theory, including SM quarks and leptons.

{\bf Higgs Vaccuum Expectation Value:} It was long-ago shown by B.W. Lee, K. Symanzik, A. Vassiliev, C. Itzykson and J-B. Zuber \cite{Lee1970,Symanzik1970a,Symanzik1970b,Vassiliev1970,ItzyksonZuber1980} that {\em the Higgs VEV $\HVEV$
receives no UVQD to all orders in perturbation theory}.  This is crucial to our argument that there is no
Higgs Fine Tuning Problem and since, \cite{Lee1970} is not widely available either in print or electronically,
an outline of  Lee's BPHZ proof is included in \cite{LynnStarkman2013a}.  C. Itzykson and J-B. Zuber prove it  in more modern functional integral language in  \cite{ItzyksonZuber1980} equation 11-176. We do not therefore dress $\HVEV$ with the subscript ``$Bare$". Since throughout this paper we shall be interested only in UVQD, and not in logarithmic divergences or finite corrections,
we make no distinction between the bare and renormalized values of $\HVEV$.

We define the physical  particle $h$ so that it cannot simply disappear into the vacuum
\begin{eqnarray}
H \equiv h + \HVEV \nonumber
\\ \langle h\rangle=0
\end{eqnarray}
and loosely refer to the real scalar  $h$  as the physical ``Higgs.'' 
Rewriting $V^{Bare}_{\LSMmth} $ in terms of $h$ and ${\vec \pi}$, 
making use of (\ref{eqn:O4Fields}), 
\begin{eqnarray}
\label{eqn:yetanotherVbare}
V^{Bare}_{\LSMmth}  &=& \frac{\lambda^2}{4}\left(h^2 + {\vec \pi}^2\right)^2 + \frac{\lambda^2}{2}\HVEV h \left(h^2 + {\vec \pi}^2\right)  \\
&+& \frac{1}{2}h^2\left(2\lambda^2\HVEV^2 + \mpisq\right)
   + \frac{1}{2}\mpisq{\vec\pi}^2  \,.\nonumber
\end{eqnarray}
We now see more clearly that $\mpisq$ is what the notation implies -- the renormalized (pole) mass-squared of the ${\vec \pi}$.
We note that $m_h^2 = \mu_{bare}^2 + 3 \lambda \HVEV^2$.
We also see the relation between the Higgs mass-squared and the pion mass-squared:
\be
\label{eq:hpi}
m_h^2 = \mpisq + 2 \lambda^2 \HVEV^2 .
\ee
{\em This relation is central to this paper because, since
$\HVEV$ has no UVQD in (\ref{eq:hpi}),  any  UVQD of $m_h^2$ must be identical to the UVQD of $\mpisq$.}

We also see in (\ref{eqn:yetanotherVbare}) that $V^{Bare}_{\LSMmth} $ contains no terms linear in $h$. 
In other words, minimization of $V^{Bare}_{\LSMmth} $ in the context of the Ward identity implies the vanishing of all tadpoles at tree-level.    
Indeed, we could have arrived at precisely the same point by instead minimizing $V^{Bare}_{\LSMmth} $ as originally specified
in (\ref{eqn:LBare-LSM}), while insisting on the vanishing of all tadpoles at tree level.

\begin{figure}[htpb]
\includegraphics[width=0.4\textwidth,height=0.3\textheight]{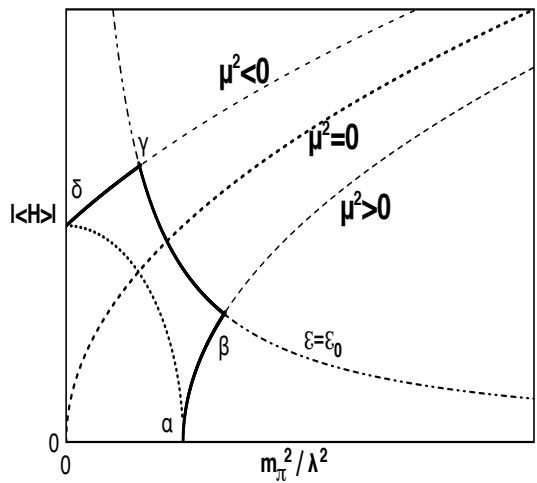}
\caption{$\vert\HVEV\vert$ vs. $m^2_\pi/\lambda^2$ quarter-plane; $\mu^2 \equiv \mpisq-\lambda^2\HVEV^2$ and $\epsilon \equiv \HVEV\mpisq$.  
 O(4) symmetry is explicitly broken off the axes; i.e., $\epsilon \neq 0$.
Axial vector current conservation is restored on both the x-axis (Wigner mode) and the y-axis (Goldstone mode).
On the x-axis  the vacuum also respects the O(4) symmetry.  On the y-axis the vacuum breaks the O(4) symmetry spontaneously.
Two paths through the plane from Wigner Mode to Goldstone Mode are shown in bold: one smooth and analytic in $\mu^2$ and $\epsilon$  (i.e. $\alpha \delta$) and one not (i.e. $\alpha \beta \gamma \delta$). This famous diagram, more than forty years old, appears in Lee \cite{Lee1970} and the textbook by C. Itzykson and J-B. Zuber  \cite{ItzyksonZuber1980}. The points ($\alpha, \beta, \gamma, \delta$) are labeled to match the latter's Figures 11-14 and 12-12. This diagram (introduced at O(4) \LSM~ tree-level) remains unchanged upon inclusion of all-loop-order UVQD corrections: in the O(4) \LSM~ with 4 real scalars; in the $SU(2)_L$ \LSM~ with a complex Higgs doublet and SM quarks and leptons.}
\label{fig6}
\end{figure}

B.W.Lee and K. Symanzik proceed to analyze the generic set of   O(4) \LSM~quantum field theories in the   $\vert\HVEV\vert$ vs. $\mpisq/\lambda^2$  quarter-plane (cf. Figure 1) for all possible values of $\mu^2$ and $\epsilon$ with
  \begin{eqnarray}
\label{eqn:MuEpsilonInMpiHVEV}
\mu^2&\equiv&\mpisq-\lambda^2\HVEV^2
\\ \epsilon &\equiv& \HVEV\mpisq
\end{eqnarray}
    Off of the axes, the value of the explicit symmetry breaking parameter $\epsilon$ is nonzero.  Lines of constant $\mu^2$ are parabolae as shown; and lines of constant $\epsilon$ are hyperbolae.
 If we start from a point on a line ($\gamma$) with $\mu^2 < 0$ and let $\epsilon \rightarrow 0$, while keeping $\mu^2$ fixed,
we will reach a point ($\delta$) on the $y$-axis where $\HVEV \neq 0, m_\pi^2 = 0$ (Goldstone mode).  On the other hand, if we start from a point ($\beta$) on the line
with $\mu^2 > 0$ and let $\epsilon \rightarrow 0$, $\mu^2$ being kept fixed, we will reach a point ($\alpha$) on the $x$-axis on which
$\HVEV = 0, m_\pi^2 = \mu^2 \neq 0$ (Wigner mode).  We note that this figure is a tree-level diagram, which we prove below remains unchanged by the inclusion of all UVQDs in the theory; 
the lines would shift due to logarithmic divergences and finite terms which we ignore throughout.
In Section 2F we will add UVQD from Standard Model fermions to this picture: it still will remain unchanged.

It is worth making several observations already at this point:
\begin{itemize}
\item  Following \GMLfull \cite{GellMannLevy1960}, Lee \cite{Lee1970}, Symanzik \cite{Symanzik1970a,Symanzik1970b}  and Itzyson and Zuber (\cite{ItzyksonZuber1980}, pg. 541), we have taken what appears to be the radical step of including an explicit symmetry breaking PCAC term in the bare Lagrangian, in order to eventually analyze the spontaneously broken theory.    
\item As K. Symanzik remarked \cite{SymanzikPC} in a private letter to R. Stora:
\begin{quote}
{\it ``Whether you like it or not, 
you have to include in the Lagrangian all possible terms consistent with locality and power counting, 
unless otherwise constrained by Ward identities."}
\end{quote}
We thus have no choice whether or not to renormalize the symmetry breaking term (\ref{eqn:LPCAC})  -- unforbidden by Ward-Takahashi identities,
it is induced anew, and receives UVQD corrections, at each order in quantum loops. 
\item Given {\em that}  (i.e. an explicit symmetry breaking term $\epsilon H$ is induced order by order in quantum loops), in order to have a spontaneously broken symmetry, and by the Goldstone theorem have massless Nambu-Goldstone Bosons,  
we will need to follow Lee \cite{Lee1970}, Symanzik \cite{Symanzik1970a,Symanzik1970b} and Itzykson and Zuber \cite{ItzyksonZuber1980}  and explicitly impose the vanishing of the explicit breaking (to each loop-order) in subsections \ref{sbscn:1loopUVQD}, \ref{sbscn:all-loop-noHiggs-FTP} and \ref{sbscn:withSMQL-noHiggs-FTP} below.
There are two ways to do so.  
     \begin{itemize}
     \item We could let $\HVEV\to0$, but this recovers the unbroken Wigner mode of the theory.  
     \item We must
     instead let $\mpisq\to0$, keeping $\HVEV\neq0$: that recovers the spontaneously broken Goldstone Mode of the theory.  
     We call this renormalization prescription, first understood by Lee and Symanzik,  the {\bf Goldstone Symmetry Restoration Condition} (Goldstone-SRC).
     \end{itemize}
\item The theory has three phases:  Wigner mode, defined by $\HVEV=0$, $\mpisq/\lambda^2>0$ (the x-axis in figure \ref{fig6}) in which the action and the vacuum are O(4) symmetric;
Goldstone mode, defined by $\mpisq/\lambda^2=0$, $\HVEV\neq0$ (the y-axis in figure \ref{fig6}), in which the action is O(4) symmetric, but the theory is spontaneously broken;
and broken phase, comprising  the rest of the $\vert\HVEV\vert-\mpisq/\lambda^2$ quarter-plane, in which the theory is explicitly broken.    Lee \cite{Lee1970} further divides the broken phase into
a Goldstone phase ($\mu^2 \leq 0$ and $\mpisq \leq \lambda^2\HVEV^2$) and a Wigner phase ($\mu^2 \geq 0$ and $\mpisq \geq \lambda^2\HVEV^2$).      The origin of the quarter-plane ($\mu^2 = \mpisq = \lambda^2\HVEV^2=0$) has conformal symmetry and may have special
properties, which we do not discuss further.

\item We see quite clearly that the physical phase of the theory is {\em not} determined by the sign of $\mu^2$.   The misconception that the sign of $\mu^2$ is
the crucial quantity for the {\em un-gauged} O(4)\LSM~ arises from the mistaken belief, common in the modern literature (eg. \cite{Peskin1995})  
that (in parameter space) one can only pass from the Wigner mode to the Goldstone mode by moving from ($\alpha$) along the x-axis of Figure \ref{fig6}, {\em through the origin}, and onto and along the y-axis to ($\delta$).
The problem with that path is that the origin, $\HVEV=0$, $\mpisq/\lambda^2=0$, is not analytic in $\mu^2$: see the textbook by  C. Itzykson and J-B. Zuber  \cite{ItzyksonZuber1980} (especially Figure 11-14), subsection \ref{sbscn:1loop-noHiggs-FTP} and \cite{Lee1970,LynnStarkman2013a,Symanzik1970a,Symanzik1970b}  for more detail.
\item Lee, Symanzik, Itzykson and Zuber all argue strongly that  {\em when renormalizing the \GML \LSM, one must consider the theory in 
the full $\vert\HVEV\vert$ vs. $\mpisq/\lambda^2$ quarter-plane}:  the Goldstone mode is then only a boundary of the parameter space of the theory that can be reached,
from the other Wigner mode boundary, by many possible paths through the quarter-plane, without passing through the origin: see Ref.  \cite{LynnStarkman2013a} and subsection \ref{sbscn:1loop-noHiggs-FTP} for more detail.
\item The Ward-Takahashi  identity linking (\ref{eqn:DivergenceofAmu}) to (\ref{eqn:epsilonequality}) shows that conservation of the axial-vector current is restored either when $\HVEV = 0$ (the x-axis Wigner mode of Figure \ref{fig6})
or $m_\pi^2 = 0$ (the y-axis Goldstone mode of Figure \ref{fig6}).  
Yet, as emphasized by C. Itzykson and J.B. Zuber \cite{ItzyksonZuber1980} and \cite{LynnStarkman2013a}, when considering the renormalization disposition of those UVQD arising in the \LSM, it is incorrect to take a path from one of these to the other that
passes through the origin, which is not analytic in $\mu^2$: one must instead take a path (e.g. either $\alpha \delta$ or $\alpha \beta \gamma \delta$) in Figure \ref{fig6} through the part of the plane that is off the axes, and avoid the origin.  From the
interior of the plane one can then move (without losing analyticity in $\mu^2$) to the Goldstone mode limit with $m_\pi^2 = 0$. 

\end{itemize}

We re-emphasize that most of the results so far, like many of the results in this paper, are rediscoveries
of the original results of the 1970s, as described for example in \cite{Lee1970,Symanzik1970a,Symanzik1970b,Vassiliev1970,ItzyksonZuber1980}.  We have taken the
small step of removing them from the explicit context of pion physics.  So for example the pion decay
constant $f_\pi$ makes no appearance here, replaced, courtesy of a Ward identity, by $\HVEV$.  
 In what follows below, we continue to make contact as far
as possible with the earlier work.

\subsection{Inclusion of 1-loop  O(4) ultra-violet quadratic divergences (UVQD)}
\label{sbscn:1loopUVQD}

We next consider the renormalization of the theory described by the bare Lagrangian (\ref{eqn:LBare-LSM}).
As noted above, we are interested only in UVQD, and will ignore (as un-interesting for this paper) logarithmic divergences and finite contributions.

{\bf Higgs VEV:} As remarked, it was long-ago shown by Lee, Symanzik\cite{Lee1970,Symanzik1970a,Symanzik1970b,Vassiliev1970} , and in Itzykson and Zuber  \cite{ItzyksonZuber1980} that {\em the Higgs VEV $\HVEV$
receives no UVQD to all orders in perturbation theory}.  This is crucial to our argument that there is no
Higgs Fine Tuning Problem.

{\bf Ward-Takahashi identity:} In its toy U(1) \LSM~ with PCAC, \cite{LynnStarkman2013a} proves that the disposition of all 1-loop UVQD corrections to the dimension-1 relevant symmetry breaking operator (i.e. the U(1) version of  (\ref{eqn:LPCAC})),
is controlled by the Ward-Takahashi identity governing the 2-pion connected Greens function, and that this 
determines (for UVQD purposes) the strength of the renormalized  symmetry breaking term. 

B.W. Lee proved \cite{Lee1970} that the Ward-Takahashi identity ((\ref{eqn:WTidentity})  and (\ref{eqn:WTdivergence})), which governs the 2-pion connnected Greens function and sets the strength of the PCAC 
relation, and the PCAC relation itself (\ref{eqn:DivergenceofAmu}) receive no UVQD corrections to all-loop-orders of perturbation theory.
Using the O(4) Ward-Takahashi identity  \cite{Lee1970,Symanzik1970a,Symanzik1970b,Vassiliev1970,ItzyksonZuber1980}, the UVQD corrections to (\ref{eqn:LPCAC}) can therefore be included in the renormalized explicit symmetry breaking term
\be
L^{1-loop; PCAC}_{\LSMmth} = \epsilon^{1-loop}_{\LSMmth} H\,
\ee
\begin{eqnarray}
\label{eqn:LSymmetryBreaking_1loop}
\epsilon^{1-loop}_{\LSMmth}  = \HVEV\mpisq
\end{eqnarray}
Here 
$\mpisq$ is the physical renormalized pion (pole) mass for three degenerate pions,
and includes {\em exactly the 1-loop UVQD contributions from virtual scalars  in  (\ref{eqn:C-Unrenorm-1loop}), (\ref{eqn:MuSquared}), (\ref{eqn:mpisqdefn}) and (\ref{eq:hpioneloop}) below} (neglecting logarithmic divergences and other finite contributions).

{\bf Feynman diagrams:} The UVQD 1-loop  Lagrangian corresponding to the bare Lagrangian (\ref{eqn:LBare-LSM}), 
evaluated at zero momentum, including all 1-loop 2-point self-energy (Figures \ref{fig1}, \ref{fig2} and \ref{fig3}) and 1-loop 1-point tadpole-function 
(Figure \ref{fig4}) UVQDs, is (cf. Appendix A):
\be
\label{eqn:L-1loop-Lsq-LSM}
L^{1-loop;\Lambda^2}_{\LSMmth} \!\!\!= C^{Unrenorm;1-loop;\Lambda^2}_{\LSMmth}\Lambda^2
\! \left(\!\!\Phi^\dagger\Phi \!-\! \frac{\HVEV^2}{2}\right) \,.
\ee
Expansion around the Higgs VEV $\HVEV$ gives
\be
\label{eqn:ExpandPhiDaggerPhi}
\Phi^\dagger\Phi - \frac{\HVEV^2}{2}=\frac{1}{2}h^2+\frac{1}{2}\pi_{3}^2+\pi_{+}\pi_{-}+\HVEV h,
\ee
while $C^{Unrenorm;1-loop;\Lambda^2}_{\LSMmth}$ is a dimensionless constant. 
The form of (\ref{eqn:L-1loop-Lsq-LSM}), explicitly calculated in Appendix A, follows from Lee and Symanzik's proof \cite{Lee1970,Symanzik1970a,Symanzik1970b,Vassiliev1970,Symanzik1969,Lee1969,Bogoliubov1957,Hepp1966,ZimmermanDeser}
 that the theory is properly renormalized throughout the  $\vert\HVEV\vert$ vs. $\mpisq/\lambda^2$  quarter-plane
{\em  with  the same UVQD graphs and counter-terms} as the symmetric Wigner-mode limit: $\HVEV^2\to0$ holding $\mpisq\neq0$. 
 Noticing that at tree level the ÒHiggsÓ mass $m_h^2\equiv2\lambda^2\HVEV^2$, 
 we  suggestively (and recognizably \cite{Veltman1981}
 \footnote{We agree with M.J.G. Veltman's \cite{Veltman1981}
  overall 1-loop Standard Model UVQD coefficient \cite{Veltman1981} in the appropriate zero-gauge-coupling limit,
but his expression  (6.1)  for the relevant (dimension 1 and 2) operators with 1-loop UVQD coefficients arising from virtual scalars and fermions in the $SU(2)_L$ \LSM~ 
would read (i.e. if taken at face value and written in the language of this paper):
$L^{1-loop;\Lambda^2}_{SM,\Phi Sector} \propto  \left(-\half h^2 +\half\pi_{3}^2 + \pi_+\pi_- - \HVEV h\right)\Lambda^2$.
This is not $SU(2)_L$  invariant before SSB  because of the relative minus sign (with which we disagree) 
between the Higgs self-energy 2-point operator and that of the Nambu-Goldstone Bosons. 
In contrast, our analogous expression is manifestly $SU(2)_L$ invariant before SSB, as it must be. 
 }, \cite{Lynn1982b,Stuart1985}) 
 rewrite   (\ref{eqn:app-quote1})  (see Appendix A):
 \begin{eqnarray}
 \label{eqn:C-Unrenorm-1loop}
C^{Unrenorm;1-loop;\Lambda^2}_{\LSMmth}\Lambda^2 &=& \left( -6\lambda^2\right)\left(\frac{\Lambda^2}{16\pi^2}\right) \\
&\text{``=''}&  -\left(\frac{3m_h^2}{16\pi^2\HVEV^2}\right)\Lambda^2\nonumber
 \end{eqnarray}
 (Here ``='' indicates that this is currently true only at tree level, although we shall find below that $m_h^2=2\lambda^2 \HVEV^2$  holds to all orders in loops for the Goldstone mode.)

The reader is reminded to take the Wigner-mode limit carefully, taking $\HVEV^2\to0$ but holding $\lambda^2$ fixed.
Appendix D shows that  (\ref{eqn:L-1loop-Lsq-LSM})  is independent of whether n-dimensional regularization or UV cutoff regularization is used. 
To fully exacerbate and reveal any Higgs-FTP, we imagine  $\Lambda\simeq M_{Planck}$ near the Planck scale.

{\bf Effective Lagrangian:} Using the bare Lagrangian $L^{Bare}_{\LSMmth}$, 
we form a 1-loop-UVQD-improved  effective Lagrangian, 
which includes all scalar 2-point self-energy and 1-point tadpole 1-loop UVQD
(but ignores 1-loop logarithmically divergent, finite contributions and vacuum energy/bubbles):
 \begin{eqnarray}
 \label{eqn:Leff-1loop-Lsq-LSM}
L^{Effective;1-loop;\Lambda^2}_{\LSMmth} &\equiv& L^{Bare;\Lambda^2}_{\LSMmth} + L^{1-loop;\Lambda^2}_{\LSMmth} \\
 &=& -\vert\partial_\mu\Phi\vert^2 - V^{Renorm;1-loop;\Lambda^2}_{\LSMmth}, \nonumber
  \end{eqnarray}
  where
 \begin{eqnarray}
 \label{eqn:V-Renorm-1loop-Lambda2}
 &&V^{Renorm;1-loop;\Lambda^2}_{\LSMmth} \!\! = \lambda^2\left[\Phi^\dagger\Phi \!\!- \!\! \frac{\HVEV^2}{2}\right]^2 - \epsilon^{1-loop}_{\LSMmth} H \nonumber\\
 &\phantom. &\quad\quad\,\,\, + \bigl(\mu^2 + \lambda^2 \HVEV^2\bigr) \left[\Phi^\dagger\Phi \!\!- \!\!\frac{\HVEV^2}{2}\right]\,
 \end{eqnarray}
The crucial observation is that the renormalized quantities $\epsilon^{1-loop}_{\LSMmth}$ and
 \be
 \label{eqn:MuSquared}
 \mu^2 \equiv \mu_{Bare}^2 -C^{Unrenorm;1-loop;\Lambda^2}_{\LSMmth}\Lambda^2
 \ee
 include all 1-loop UVQD in the renormalized effective potential (\ref{eqn:V-Renorm-1loop-Lambda2}).

Using (\ref{eqn:ExpandPhiDaggerPhi}) we identify the common coefficient of the quadratic terms in $\pi_i$ as
 \be
\label{eqn:mpisqdefn}
m_\pi^2 = \mu^2 + \lambda^2 \HVEV^2 .
\ee
Similarly
\be
m_h^2 = \mu^2 + 3 \lambda^2 \HVEV^2
\ee
so that
\be
\label{eq:hpioneloop}
m_h^2 = m_\pi^2 + 2 \lambda^2 \HVEV^2 .
\ee

Equations (\ref{eqn:MuSquared}, \ref{eqn:mpisqdefn} and \ref{eq:hpioneloop}) are the 1-loop UVQD-inclusive versions of (\ref{eqn:HVEVsq} and \ref{eq:hpi}).
We see that UVQD at one loop do not disturb the tree-level relations;  we will see that this fact remains
true to all orders in loop perturbation theory for UVQD as well in Section II E.

It is instructive to eliminate the un-physical parameter $\mu^2$ and rewrite the 1-loop renormalized potential in terms of three physical observables $\lambda^2 , \mpisq$ and $\HVEV$:
\begin{eqnarray}
 \label{eqn:V-Renorm-1loop-Lambda2_b}
 &&V^{Renorm;1-loop;\Lambda^2}_{\LSMmth}  \nonumber\\
 &\phantom. &\quad= \lambda^2\left[\frac{h^2}{2} + \frac{\pi_3^2}{2} + \pi_+\pi_- + \HVEV h\right]^2 \!\! \\ 
&\phantom. &\quad+ \mpisq \left[\frac{h^2}{2} + \frac{\pi_3^2}{2} + \pi_+\pi_- \right]  +  \!\!\left( \HVEV\mpisq -\epsilon^{1-loop}_{\LSMmth}\right) h \,,\nonumber
  \end{eqnarray}
Because of the way it enters (\ref{eqn:V-Renorm-1loop-Lambda2_b}), $\mpisq$ 
{\em is} the physical renormalized pseudo-scalar pion (pole) mass-squared entering into the divergence of the axial-vector current in  (\ref{eqn:WTdivergence}) and (\ref{eqn:DivergenceofAmu}).

Equation (\ref{eqn:V-Renorm-1loop-Lambda2_b}) appears to retain terms linear in $h$, i.e. tadpole terms.  This leads
Lee and Itzykson and Zuber to impose a Higgs Vacuum Stability Condition (our name) essentially by hand \cite{Lee1970,ItzyksonZuber1980}:
\smallskip

{\bf Higgs-VSC: The physical Higgs   particle must not disappear into the exact UV-corrected vacuum:} 
 \begin{enumerate}
 \item For the purposes of this paper, we take the UV-corrected vacuum to be the UVQD-corrected vacuum.
\item The UVQD-corrected vacuum includes all perturbative UVQD corrections, including
 	\begin{itemize}
	\item 1-loop  when referenced in subsections \ref{sbscn:1loopUVQD}, \ref{sbscn:Higgs-FTP} and \ref{sbscn:1loop-noHiggs-FTP},
	 \item 1PI multi-loop when referenced in subsection \ref{sbscn:all-loop-noHiggs-FTP} and Appendix B
	 \item 1-loop $SU(2)_L$ \LSM~ with SM quarks and leptons in subsection  \ref{sbscn:withSMQL-noHiggs-FTP} and Appendix C.
 	 \end{itemize}
\item Exact tadpole renormalization is to be imposed to all orders in perturbation theory. 
\item {\bf Stationary Condition:} By definition,  $H=\HVEV+h$, and $\langle\Phi^\dagger\rangle\langle\Phi\rangle=\half\HVEV^2$ is the exact square of the VEV.
The physical Higgs particle $h$ must have exactly zero VEV in order that it not simply disappear into the vacuum: $\langle h\rangle\equiv0$. It follows that {\em the theory is required to be stationary at the minimum of the renormalized potential, i.e. the vacuum at $H=\HVEV$ and $\langle \vec \pi \rangle =0$}.
\item It is important to realize that Higgs-VSC tadpole renormalization does not constitute fine-tuning; 
 rather it is a stability condition on the vacuum and excited states of the theory.
 \end{enumerate}

 However, {\em at least for 1-loop UVQD, in the O(4) \LSM~  it is not necessary to impose the Higgs-VSC, because it is automatically 
 enforced by the  Ward Takahashi identity.}  In other words, extending the results in \cite{LynnStarkman2013a} to O(4) and imposing 
 (\ref{eqn:LSymmetryBreaking_1loop}) 
 on (\ref{eqn:V-Renorm-1loop-Lambda2_b}), we can write 
 \begin{eqnarray}
 \label{eqn:V-Renorm-1loop-Lambda2_c}
 &&V^{Renorm;1-loop;\Lambda^2}_{\LSMmth}  \nonumber\\
 &\phantom. &\quad= \lambda^2\left[\frac{h^2}{2} + \frac{\pi_3^2}{2} + \pi_+\pi_- + \HVEV h\right]^2 \!\! \\ 
&\phantom. &\quad+ \mpisq \left[\frac{h^2}{2} + \frac{\pi_3^2}{2} + \pi_+\pi_- \right]   \nonumber
  \end{eqnarray}
with no tadpole terms. This is true at one-loop, just as it was in the tree-level theory.  
  This fact appears to have heretofore been unappreciated.  
  Thus Lee \cite{Lee1970} quite reasonably  imposed the Higgs-VSC as an external condition on the theory by enforcing $f_\pi=\HVEV$ everywhere in what
  was for him the 
 $f_\pi-\mpisq/\lambda^2$ quarter-plane
 \footnote{ 
This is distinct from the observation that in the Goldstone mode this Higgs-VSC condition
 is also guaranteed by the Goldstone-SRC, i.e. by $\mpisq=0$. }.

{\bf Stable vacuum:} Along any path through the quarter-plane of Figure \ref{fig6}, either smooth (e.g.  $\alpha \delta$) or not (e.g.  $\alpha \beta \gamma \delta$), from the Wigner mode to the Goldstone mode, Ward-Takahashi identities automatically enforce the Higgs-VSC. The physical Higgs particle is prevented from simply disappearing into the vacuum. The vacuum is always stable (i.e. the theory is stationary, as we required above) at the minimum of the effective potential $V^{Renorm;1-loop;\Lambda^2}_{\LSMmth}$ (Equation (\ref{eq:form1loop}) below), which is at $H=\HVEV$ and $\vec \pi =0$.

\begin{figure}[htpb]
\includegraphics[width=0.4\textwidth,height=0.3\textheight]{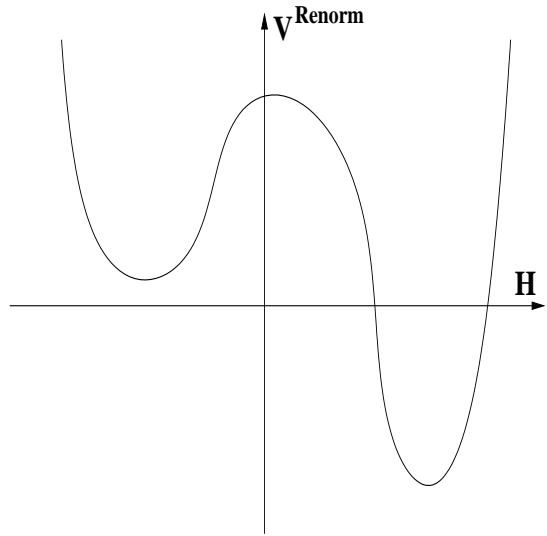}
\caption{Cross section of the \GMLfull potential $V^{{\rm Renorm;All-loops},\Lambda^2}_{\LSMmth}  = V^{{\rm Tree}}_{\LSMmth}  = V^{{\rm Renorm;All-loops},\Lambda^2}_{\LSMmth +SM q \& l; \Phi}$ 
 through its deepest point  (the vacuum $H=\HVEV ;\langle \vec \pi \rangle =0$) and its local maximum,  
for $\mu^2<0$, and fixed $\mpisq>0$.   When $\mpisq=0$, the minima
are degenerate, and $V$ acquires ``bottom of the wine-bottle" symmetry, together with Nambu Goldstone bosons and their $SU(2)_{L-R}$ chiral symmetry. J.C. Taylor draws a 3-dimensional version (\cite{JCTaylor1976} pg. 31 Figure 5.2) of this old and famous potential in his chapter on soft pions.}
\label{fig5}
\end{figure}

The 1-loop UVQD-improved renormalized potential (\ref{eqn:V-Renorm-1loop-Lambda2_c}) can be re-written so that it looks just  like the tree-level result (\ref{eq:form}):
\begin{eqnarray}
\label{eq:form1loop}
V^{Renorm;1-loop;\Lambda^2}_{\LSMmth}  &=& \lambda^2\left[ \Phi^\dagger\Phi - \frac{1}{2}\left(\HVEV^2  -  \frac{\mpisq}{\lambda^2}\right) \right]^2 \nonumber
\\ &\phantom{-}& \quad -\HVEV \mpisq H
\\ &=& V^{\Tree}_{\LSMmth} \nonumber
\end{eqnarray}
In  (\ref{eq:form1loop}), 
$\mpisq$
is {\em still} the 
physical renormalized pseudo-scalar pion (pole) mass-squared of (\ref{eqn:mpisqdefn}),
and 
\be
m_h^2 \equiv \mpisq + 2\lambda^2 \HVEV^2 \geq \mpisq.
\ee
A cross-section of $V^{Renorm;1-loop;\Lambda^2}_{\LSMmth}$, with $ \mu^2 = \mpisq-\lambda^2 \HVEV^2<0$ and $\mpisq>0$ is plotted in Figure \ref{fig5}.

{\bf ``New" 1-loop UVQD corrections to the \GMLfull PCAC relation:} Following \cite{LynnStarkman2013a}, we overlay equations (\ref{eqn:LSymmetryBreaking_1loop}), (\ref{eqn:C-Unrenorm-1loop}), 
(\ref{eqn:MuSquared}) and (\ref{eqn:mpisqdefn}) to reveal that the divergence of the axial-vector current does indeed receive 1-loop UVQD corrections:
\begin{eqnarray}
\label{PCAC1-loopUVQD}
\partial_{\mu} {\vec J}^{5}_{\mu} = &-& \HVEV \mpisq \vec \pi
\\ = &-&\HVEV [ \mu_{Bare}^2 + \lambda^2 \HVEV^2  
\\ &\phantom{-}& \quad - C^{Unrenorm;1-loop;\Lambda^2}_{\LSMmth}\Lambda^2 ] ~ \vec \pi \,. \nonumber
\end{eqnarray}
These ``new" UVQD arise in the \GMLfull ($\epsilon \neq 0$) \LSM, but not in the Schwinger ($\epsilon \equiv 0$) O(4)  \LSM~ \cite{Schwinger1957}. {\em But Ward-Takahashi identities ensure that the UVQD corrections to the PCAC relation (i.e. those in  (\ref{eqn:C-Unrenorm-1loop})) are the same as the UVQD corrections to the effective potential $V^{Renorm;1-loop;\Lambda^2}_{\LSMmth}$ in (\ref{eq:form1loop})}. Our 1-loop UVQD corrections to the PCAC relation  do not appear in the most modern textbooks \cite{Peskin1995,Kaku1993,Weinberg1995} and may be widely unfamiliar to modern audiences.

{\bf Explicit breaking of $SU(2)_{L-R}$ Nambu Goldstone boson  chiral symmetry in the \GMLfull model:} Following \cite{LynnStarkman2013a}, we re-write (\ref{eq:form1loop})  in the unitary Kibble representation of  $\Phi$ \cite{Ramond2004,Georgi2009} with transformed fields $\tilde H$ and $\vec{\tilde \pi}$, with VEVs $\langle \tilde H\rangle =\HVEV$ and $\langle {\vec{\tilde\pi}} \rangle =0$ and
\begin{eqnarray}
\Phi_{SU(2)_L \times SU(2)_R} &=& \frac{1}{\sqrt{2}} {\tilde H} U \nonumber
\\ U &\equiv& \exp \left[ i \frac {{\vec \sigma} \cdot \vec{\tilde \pi}}{\HVEV} \right]
\end{eqnarray}

The kinetic term
\be
 \label{eqn:LKinetic-KibbleRep}
 L^{Kinetic}_{\LSMmth} = -\frac{1}{2} \left( \partial_{\mu} {\tilde H} \right)^2 - \frac{1}{4} {\tilde H}^2 Tr \left[ \partial_{\mu} U^{\dagger} \partial_{\mu} U \right] 
 \ee
is invariant under $SU(2)_L \times SU(2)_R$ transformations, and  regards the $ {\vec{\tilde\pi}}$ as true Nambu Goldstone bosons, with only derivative couplings  \cite{JCTaylor1976,Georgi2009}.

The renormalized potential
\begin{eqnarray}
\label{eqn:V1-loopUnitaryRep}
V^{Renorm;1-loop;\Lambda^2}_{\LSMmth}  &\!=\!& \frac{1}{4} \lambda^2\left[ {\tilde H}^2 \!-\! \left(\HVEV^2  \!-\!  \frac{\mpisq}{\lambda^2}\right) \right]^2
\\ &\phantom{-}& \quad - \HVEV\mpisq \tilde H \cos{\frac{\sqrt{{\vec{\tilde \pi}} \cdot {\vec{\tilde \pi}}}}{\HVEV}}   \nonumber
\end{eqnarray} 
preserves (with constant ${\vec \alpha} \equiv {\hat \alpha} \alpha$)  ordinary homogenous $SU(2)_{L+R}$ symmetry 
\begin{eqnarray}
\label{SU2Vector}
{{\tilde H}} &\rightarrow& {{\tilde H}} 
\\ U &\rightarrow& \exp \left[ \frac{i}{2} {{\vec \sigma} \cdot \vec{\alpha}} \right] \cdot U \cdot \exp \left[ - \frac{i}{2} {{\vec \sigma} \cdot \vec{\alpha}} \right] \,\nonumber
\\ {\vec{\tilde \pi}} \cdot {\vec{\tilde \pi}} &\to& {\vec{\tilde \pi}} \cdot {\vec{\tilde \pi}}
\\ {\vec{\tilde \pi}} &\to& {\vec{\tilde \pi}} \cos{\alpha} + {\vec{\tilde \pi}} \times {\hat \alpha} \sin{\alpha} +2{\hat \alpha} \left[ {\hat \alpha} \cdot {\vec{\tilde \pi}} \right] { \sin^2{ \frac{\alpha}{2} } } \nonumber
\end{eqnarray}
and CVC, so that 
$\partial_{\mu} {\vec J}_{\mu} = 0$.

Now focus on the $SU(2)_{L-R}$ subgroup of pure chiral transformations (with constant ${\vec \theta}$):
\begin{eqnarray}
\label{InhomogenousShiftSymmetry}
{{\tilde H}} &\rightarrow& {{\tilde H}}
 \\ U &\rightarrow& exp \left[ \frac{i}{2} {{\vec \sigma} \cdot \vec{\theta}} \right] \cdot U \cdot exp \left[ \frac{i}{2} {{\vec \sigma} \cdot \vec{\theta}} \right] \,.\nonumber
\end{eqnarray}
This is sometimes called ``inhomogenous $SU(2)_{L-R}$ Nambu Goldstone shift symmetry" because the NGB are (to lowest order) shifted by constants
\be
{\vec{\tilde\pi}} \rightarrow {\vec{\tilde\pi}} + {\HVEV {\vec \theta}} ~ + {\cal O}(\theta^2) \,.
\ee

The PCAC relation
\begin{eqnarray}
\label{eqn:PCACUnitaryRep}
 \partial_{\mu} {\vec J}_{\mu}^5 &=& -\HVEV\mpisq \tilde H \frac{\vec{\tilde \pi}}{ \sqrt{   \vec{\tilde \pi}    \cdot \vec{\tilde \pi}    } } \sin{\frac{\sqrt{{\vec{\tilde \pi}} \cdot {\vec{\tilde \pi}}}}{\HVEV}}
\end{eqnarray}
is driven by the explicit symmetry breaking $\mpisq \neq 0$ term in
 (\ref{eqn:V1-loopUnitaryRep}), and breaks explicitly the inhomogeneous NGB  $SU(2)_{L-R}$ chiral symmetry obeyed by the kinetic term (\ref{eqn:LKinetic-KibbleRep}) in the bare Lagrangian. 
{\em It is therefore not possible to change the results of this paper
by re-scaling $\langle {\tilde H} \rangle$} \cite{LynnStarkman2013a}
\be
\langle {\tilde H} \rangle^2 \to \HVEV ^2 -\frac{\mpisq}{\lambda ^2}
\ee
{\em in  (\ref{eqn:V1-loopUnitaryRep}): such re-scaling would also violate the theory's stationary condition $H \equiv h+\HVEV, \langle h\rangle =0$} \cite{LynnStarkman2013a}.

Most of the basic results in this subsection  are not new \cite{Lee1970,Symanzik1970a,Symanzik1970b,Vassiliev1970,Symanzik1969,Gervais1969,Lee1969,ItzyksonZuber1980}.
(Except:  the effective Lagrangian presentation; the absorption of all UVQD, including those to the PCAC relation,
into $\mpisq$; and the observation that Ward-Takahashi identities automatically enforce the Higgs-VSC.) 
We simply point out here that 
{\em Ward-Takahashi identities are sufficient to force all UVQD in the  S-Matrix of the O(4) \LSM~ to be absorbed 
into the physical renormalized pseudo-scalar pion (pole) mass-squared, $\mpisq$.}
We also remark that this is not surprising.  The quadratic divergences are ultra-violet constants. They therefore retain no ``memory'' of the low-energy 
($\HVEV\ll\Lambda$) breaking of the O(4) symmetry.  $H$ and $h$ are the same to UVQD, and, by the O(4) symmetry, indistinguishable from $\pi_{i}$.

\subsection{$\mpisq\neq0$: A Definition of  the Naturalness  and Higgs Fine-Tuning Problems}
\label{sbscn:Higgs-FTP}

For $\mpisq\neq0$, the 1-loop-improved O(4) \LSM~ effective  Lagrangian (\ref{eqn:Leff-1loop-Lsq-LSM}), has 1-loop UVQD contributions from  
$L^{1-loop;\Lambda^2}_{\LSMmth}$, in equations (\ref{eqn:L-1loop-Lsq-LSM}) and (\ref{eqn:C-Unrenorm-1loop}),
to both the pion and Higgs masses. 
These are cancelled by the bare term in (\ref{eqn:Leff-1loop-Lsq-LSM})
but may leave very large finite residual contributions: 

\begin{eqnarray}
\label{eqn:mpisq_rexpressed}
\mpisq &=&   \mu_{bare}^2  + \left(6\lambda^2\right)\frac{\Lambda^2}{16\pi^2}+ \lambda^2 \HVEV^2
\shortintertext{and}
m_h^2 &=&  \mu_{bare}^2   + \left(6\lambda^2\right)\frac{\Lambda^2}{16\pi^2}+ 3\lambda^2 \HVEV^2
\end{eqnarray}

Now imagine weak interaction experiments require  $G_\mu^{Exp} \HVEV^2 = {\cal O}(1)$,
where the weak scale is set by the {\em experimental} muon decay constant $G_\mu^{Exp}=1.16637\times10^{-5}{\rm GeV}^{-2}=(292.807{\rm GeV})^{-2}$. Since $\HVEV$  is a free parameter 
(defined by (\ref{eqn:WTidentity}) and the minimum of $V^{Renorm;1-loop;\Lambda^2}_{\LSMmth}$ in (\ref{eq:form1loop}), but otherwise not determined by the internal self-consistency of the theory), and receives no UVQD \cite{Lee1970,Symanzik1970a,Symanzik1970b,Vassiliev1970,ItzyksonZuber1980}, 
there is no problem in choosing its numerical value from experiment. 

But problems appear if experiments also require weak-scale  $G_\mu^{Exp} \mpisq = {\cal O}(1)$ and $G_\mu^{Exp}m_h^2 = {\cal O}(1)$.
For Planck scale ultra-violet cut-off $\Lambda\simeq M_{Pl}$, 
absorption of the UVQD into  $m_h^2$ and $\mpisq$ 
requires fine-tuning $\mu_{bare}^2$ to within $\left(G_\mu^{Exp} M_{Pl}^2\right)^{-1}\simeq 10^{-32}$ 
of its natural value \cite{Veltman1981,Susskind1979,Wilson,Raby2010}.
This is a ÒHiggs Fine-Tuning ProblemÓ and a violation of ÒNaturalnessÓ which has been variously defined to be the demand that:
\begin{itemize}
\item observable properties of the theory be stable against minute variations of fundamental parameters \cite{Susskind1979,Wilson}; and
\item electroweak radiative corrections be the same order (or much smaller) than the actually observed values \cite{Veltman1981}.
\end{itemize}
The problem is widely regarded as exacerbated in O(4) \LSM~ (and the SM)  because even if one fine tunes the parameters at 1-loop,
one must retune them from scratch at 2-loop, and again at each order in loops.  This is sometimes called the Technical Naturalness Problem.
For weak-scale O(4) \LSM~ with $\mpisq\neq0$ , these problems are all the same problem, the Higgs-FTP.\footnote{
Weak-scale O(4) \LSM~   also separately suffers a ``Weak-Planck Hierarchy" problem 
because it is unable to predict or explain the enormous splitting between the weak scale and the next larger scale 
(e.g. classical gravitational physics Planck scale $M_{Pl}$).  We make no claim to address that aesthetic problem here.}

O(4) \LSM~ in Wigner mode \cite{Lee1970,Symanzik1970a,Symanzik1970b,Vassiliev1970,Symanzik1969,Gervais1969,Lee1969},
with unbroken O(4) symmetry and four degenerate massive scalars (i.e. 1 scalar + 3 pseudo-scalars), 
is defined as the limit  $\epsilon=\HVEV \mpisq\to0$, holding $\mpisq\neq0$   while $\HVEV\to0$ ({\it i.e.} the x-axis in Figure \ref{fig6}). 

Including 1-loop UVQD (but ignoring logarithmic divergences, un-interesting finite contributions and vacuum energy/bubbles) 
the 1-loop-corrected effective Lagrangian is:
\be
L^{Effective;1-loop;\Lambda^2}_{\LSMmth} \!\!\xrightarrow[\HVEV=0, \mpisq\neq0]{} L^{Effective;1-loop;\Lambda^2}_{Wigner\LSMmth} 
\ee
where
\be
L^{Effective;1-loop;\Lambda^2}_{Wigner\LSMmth} = -\vert\partial_\mu\Phi\vert^2 - \lambda^2\left[\Phi^\dagger\Phi + \frac{\mpisq}{2\lambda^2}\right]^2 \,
\ee                                             
and the Higgs VEV is zero
\begin{eqnarray}
H\to h; &\quad& \HVEV \to 0
\end{eqnarray}

Since in O(4) Wigner mode
\begin{eqnarray}
\label{WignerNaturalScale}
m_h^2 = \mpisq &=& \mu_{bare}^2 + \lambda^2 \HVEV^2 + \left(6\lambda^2\right)\frac{\Lambda^2}{16\pi^2}
\end{eqnarray}
the natural scale of $m_{h}^2$ and $\mpisq$ is $\sim \Lambda^2$.
Weak-scale Wigner mode therefore suffers a Higgs-FTP, because it requires  fine-tuning the Higgs and pion masses (i.e. by fine-tuning $\mu_{Bare}^2$ against $\Lambda^2$) to achieve un-natural   ÒexperimentalÓ values, $m_{h}^2 ,\mpisq \ll \Lambda^2$. 
More generally, the theory suffers this Higgs-FTP everywhere in the $\vert\HVEV\vert$ vs. $\mpisq /\lambda^2$ quarter-plane in Figure \ref{fig6} {\em except on the y-axis} (as we show in subsection \ref{sbscn:1loop-noHiggs-FTP} below).

Note that setting $\HVEV =0$ in  (\ref{PCAC1-loopUVQD}), we see that (including all UVQD) the axial-vector current is conserved in Wigner mode. 
\be
\label{WignerCAC}
\partial_\mu\vec{J}_\mu^5 \!\!\xrightarrow[\HVEV \rightarrow 0, \mpisq\neq0]{}  0
\ee
{\em Thus, because the unbroken O(4) symmetry is restored in the Wigner mode  limit as a consequence of (\ref{WignerCAC}), the conservation of the axial-vector current there cannot be considered to be  fine-tuned} (even though $m_{h}^2$ and $\mpisq$ are).

\subsection{1-loop Goldstone Exception: Weak-scale spontaneously broken O(4) \LSM~has no Higgs-FTP at 1-loop}
\label{sbscn:1loop-noHiggs-FTP}

Goldstone-mode O(4) \LSM, with spontaneously broken O(4) symmetry, 1 massive scalar and 3 exactly mass-less pseudo-scalar Nambu-Goldstone Bosons, 
is defined at 1-loop as the other (not Wigner-mode) limit of the O(4) \LSM~ in which $\epsilon^{1-loop}_{\LSMmth} = \HVEV \mpisq\to0$  \cite{Lee1970,Symanzik1970a,Symanzik1970b,Vassiliev1970,Symanzik1969,Gervais1969,Lee1969}.   
Namely $\HVEV\neq0$ while $\mpisq\to0$ 
({\it i.e.} the y-axis in Figure \ref{fig6}).
Including 1-loop UVQD (and again ignoring un-interesting logarithmic divergences, finite contributions and vacuum energy/bubbles)
 the 1-loop-corrected effective Lagrangian is:
 \be
L^{Effective;1-loop;\Lambda^2}_{\LSMmth} \!\!\!\xrightarrow[\HVEV\neq0, \mpisq=0]{}\! L^{Effective;1-loop;\Lambda^2}_{Goldstone;\LSMmth} 
\ee
where
\be
 \label{eqn:Eff-1loop-GM-orig}
L^{Effective;1-loop;\Lambda^2}_{Goldstone;\LSMmth} = -\vert\partial_\mu\Phi\vert^2 - \lambda^2\left[\Phi^\dagger\Phi - \frac{\HVEV^2}{2}\right]^2 \, .
\ee  
Here $H =  h + \HVEV$ and
\begin{flalign}
\quad\quad&\mpisq \equiv  \mu_{Bare}^2  +\lambda^2\HVEV^2 - C^{Unrenorm;1-loop;\Lambda^2}_{\LSMmth} \Lambda^2& \nonumber
\\ \quad\quad&\quad \xrightarrow[\HVEV\neq0, \mpisq=0]{}  m^2_{\pi;NGB;\LSMmth}   \\
\quad\quad&\phantom{\mpisq} \equiv \mu_{Bare}^2 \nonumber +\lambda^2\HVEV^2 - C^{Unrenorm;1-loop;\Lambda^2}_{\LSMmth} \Lambda^2 \notag\\
\quad\quad&\quad\,\,\,\equiv 0
\end{flalign}

Insight into these equations requires carefully defining the properties of the vacuum, 
including all effects of 1-loop-induced UVQD, after spontaneous symmetry breaking (SSB).
The famous theorem of Goldstone \cite{Goldstone1961}, proved by Goldstone, Salam and Weinberg  \cite{Goldstone1962}, states that:  
``if there is a continuous symmetry transformation under which the Lagrangian is invariant, 
then either the vacuum state is also invariant under the transformation, or there must exist spinless particles of zero mass.''
The  O(4) invariance of the \LSM~ Lagrangian (\ref{eqn:LBare-LSM}) is explicitly broken by the $\epsilon H$ term everywhere in the $\vert\HVEV\vert-\mpisq/\lambda^2$ quarter-plane,
except of course for $\epsilon=0$.  In Wigner mode ($\epsilon=0$ because $\HVEV=0$, but $\mpisq\neq0$), the vacuum retains the 
O(4) invariance.  
In Goldstone mode ($\epsilon=0$ because $\mpisq=0$, but $\HVEV\neq0$), the presence of a VEV for $\Phi$ (as in (\ref{eqn:Eff-1loop-GM-orig})) spoils the invariance of the vacuum state under the O(4) symmetry,
hence, by the Goldstone Theorem, there must exist three spinless  zero mass particles, the Nambu Goldstone bosons \cite{Nambu1959,Goldstone1961,Goldstone1962} with $m^2_{\pi;NGB;\LSMmth}=0$.

As shown by Lee \cite{Lee1970}, Symanzik \cite{Symanzik1970a,Symanzik1970b} and Itzykson and Zuber \cite{ItzyksonZuber1980}, because the PCAC term breaks $SU(2){L-R}$ chiral symmetry explicitly, 
the Goldstone Theorem is not automatic order-by-order in quantum loops as the \GMLfull \LSM~ is renormalized, 
and so, {\em in order to force the theory to Goldstone mode, one must impose (by hand!) the Goldstone Symmetry Restoration Condition (Goldstone-SRC), a renormalization prescription.  
After SSB, the Goldstone Theorem must be an exact property (to all loop orders) of the Goldstone-mode O(4) \LSM~  vacuum and its excited states:}
\begin{itemize}
\item The Goldstone-SRC includes all perturbative UVQD corrections, including
	\begin{itemize}
	\item1-loop  when referenced in subsections \ref{sbscn:1loop-noHiggs-FTP} and \ref{sbscn:withSMQL-noHiggs-FTP} and Appendices A and C.
	\item1PI Multi-loop when referenced in subsection \ref{sbscn:all-loop-noHiggs-FTP} and in Appendix B.
	\end{itemize}
\item In generic  theories spanning the   $\vert\HVEV\vert$ vs. $\mpisq/\lambda^2$  quarter-plane (Figure \ref{fig6}),
	it is necessary to impose this Goldstone-SRC (essentially by hand)  to force the 1-loop UVQD-corrected theory to the Goldstone-mode O(4) \LSM~ limit, 
	{\it i.e.} onto the y-axis, $\mpisq\to m^2_{\pi;NGB;\LSMmth}\equiv0$, where $m^{2}_{\pi ;NGB;\GML}$ is the NGB mass.
\item To higher loop-orders, for spontaneously broken O(4) \LSM~,  one must explicitly enforce the Goldstone Theorem order-by-quantum-loop-order
	so  the pions remain exactly massless to all orders of loop perturbation theory. 
	That is the purpose of Lee/Symanzik's Goldstone-SRC.  {\em One thus has no choice but to operate in the $\vert\HVEV\vert-\mpisq/\lambda^2$ quarter-plane, because
	at each order in loops,  $\mpisq$ reappears in UVQD multi-loop 1PI graphs plus nested counter-terms, and at each order (or at least at the final order to which
	one calculates) one must explicitly take $\mpisq\to m^2_{\pi;NGB;\LSMmth}\equiv0$ to realize the Goldstone Theorem.}
\item We have ignored certain infra-red (IR) subtleties \cite{ItzyksonZuber1980} as beyond the scope of this paper.
\end{itemize}

{\bf Proper renormalization of the spontaneously broken O(4)\LSM:} B.W. Lee \cite{Lee1970}, K. Symanzik \cite{Symanzik1970a,Symanzik1970b} 
and Itzykson and Zuber \cite{ItzyksonZuber1980} argue strongly that
{\em one {\bf must} implement the renormalization of the theory throughout the entire $\vert\HVEV\vert$ vs. $\mpisq/\lambda^2$ quarter-plane in Figure \ref{fig6}: this is a requirement, not a choice, for proving the renormalizabiity of the spontaneously broken O(4)\LSM.}

We therefore quote extensively from the discussion, in the textbook by C. Itzykson and J-B. Zuber (\cite{ItzyksonZuber1980} pg. 548), about our Figure \ref{fig6} (their Figures 11-14 and 12-12) from their discussion of the O(N)\LSM:

\begin{quote}
{\em ``We are now in a position to discuss the"} renormalization of the {\em ``interesting case with spontaneous symmetry breaking. At first sight, it would seem reasonable to start ...  with $\mu^2 >0$ and then continue to ... $\mu^2 <0$"} along the path from  the point $\alpha$, through the origin,  to  the point $\delta$ in Figure \ref{fig6}. {\em ``The trouble is that this procedure involves a transition through a singular point, Physical quantities such as $\mpisq (\mu^2, \epsilon)$ will not be analytic in $\mu^2$ for $\epsilon=0$"} as demonstrated in their Figure 11-14.  {\em ``To cope with this difficulty we may introduce ...  $\epsilon\neq 0$ to turn around the singularity. In Figure 11-14"} and Figure 12-12 {\em ``we show how the Goldstone"} mode {\em ``may be reached along the path $\alpha \beta \gamma \delta$ by successively varying $\epsilon$ and $\mu^2$. ... it follows that counter-terms pertaining to the symmetric theory (point $\alpha$) will insure the finiteness at the"} explicitly {\em ``broken symmetry point $\beta$. Symmetric mass counter-terms will lead from $\beta$ to $\gamma$ corresponding to the variation in $\mu^2$. Finally in the limit of vanishing $\epsilon$ (point $\delta$)"}
the Fourier Transform of the inverse propagator, {\bf evaluated at zero momentum}, satisfies: 
 \be
 v_l T^{\alpha}_{lm} \langle 0\vert T\left[ \phi_m (x)\phi_k (0)\right] \vert 0\rangle^{-1}_{FT}=0
 \ee
for O(N) fields $\phi_m$, infinitesimal group generators $T^{\alpha}$ and vacuum expectation values $\langle 0\vert\phi_l\vert 0\rangle=v_l$. 
 {\em ``This means that the (N-1) transverse bosons are massless, being the"} Nambu {\em ``Goldstone bosons of the spontaneously broken symmetry."} 
\end{quote}

Many issues and subtleties are resolved in the $\vert\HVEV\vert$ vs. $\mpisq/\lambda^2$ quarter-plane. Among those discussed in \cite{LynnStarkman2013a} are the so-called ``Higgs fine-tuning discontinuity" and the requirement that (by definition) the theory be stationary at $H=h+\HVEV$, where $\HVEV$ has no UVQD and $\langle h \rangle =0$.

{\bf Goldstone mode 1-loop UVQD-improved effective Lagrangian:} It is easy to see, from (\ref{eqn:V-Renorm-1loop-Lambda2_c}) and (\ref{eq:form1loop}), that  $V^{Renorm;1-loop;\Lambda^2}_{\LSMmth}$, shown in Figure \ref{fig5} with $\mu^{2}<0$ and $\mpisq\neq0$,
only has NGB when (``bottom-of-the-wine-bottle Goldstone symmetry") is restored: i.e. in Goldstone mode with
\be
\mpisq \to m^2_{\pi;NGB;\LSMmth}\equiv0.
\ee
Re-write (\ref{eqn:Leff-1loop-Lsq-LSM}) after tadpole renormalization, 
 \be
 \label{eqn:Leff-1loop-Lsq-LSM-tadpoled}
L^{Effective;1-loop;\Lambda^2}_{Goldstone;\LSMmth} \!=\! -\vert\partial_\mu\Phi\vert^2 - V^{Renorm;1-loop;\Lambda^2}_{Goldstone;\LSMmth}, 
  \ee
  where, as in (\ref{eqn:V-Renorm-1loop-Lambda2_c}),
 \begin{eqnarray}
 \label{eqn:V-Renorm-1loop-Lambda2-GM}
 V^{Renorm;1-loop;\Lambda^2}_{Goldstone;\LSMmth}&&= \lambda^2\left[\frac{h^2}{2}  + \frac{\pi_3^2}{2} + \pi_+\pi_- + \HVEV h \right]^2 \nonumber \\
&& \!\!\!\!\! \!\!\!\!\! \!\!\!\!\! \!\!\!\!\! + m^2_{\pi;NGB;\LSMmth}\left[\frac{h^2}{2}  + \frac{\pi_3^2}{2} + \pi_+\pi_-\right].
  \end{eqnarray}

Again, as in (\ref{eqn:mpisqdefn}) and (\ref{eqn:mpisq_rexpressed}),
  a crucial observation in this paper is that {\em proper 1-loop enforcement of the Goldstone Theorem $m^2_{\pi;NGB;\LSMmth}\equiv0$,
requires imposition (essentially by hand) of Lee/SymanzikÕs  Goldstone-SRC \cite{Lee1970,Symanzik1970a,Symanzik1970b,Vassiliev1970,Symanzik1969,Gervais1969,Lee1969}
\be
0\equiv  m^2_{\pi;NGB;\LSMmth}  =  \mu_{Bare}^2 + 6\lambda^2 \frac{\Lambda^2}{16\pi^2}+\lambda^{2}\HVEV^{2}
  \ee
exactly and identically.    }
Only at that point  can we write
\begin{eqnarray}
\label{eqn:Venorm1loopGM}
 V^{Renorm;1-loop;\Lambda^2}_{Goldstone;\LSMmth}&& = \lambda^2\left[\frac{h^2}{2}  + \frac{\pi_3^2}{2} + \pi_+\pi_- + \HVEV h \right]^2  \nonumber \\
&&= \lambda^2\left[\Phi^\dagger\Phi^2 - \frac{\HVEV^2}{2} \right]^2\,. 
  \end{eqnarray}

{\bf 1-loop UVQD corrections to the \GMLfull PCAC relation, and conservation of the axial-vector current,  in Goldstone mode:} Following \cite{LynnStarkman2013a}, the 1-loop UVQD corrections to the O(4) PCAC relation also vanish (without fine-tuning) in Goldstone mode:
\begin{eqnarray}
\label{PCAC1-loopUVQDGoldstone}
\partial_{\mu} {\vec J}^{5}_{\mu} &&= -\HVEV [ \mu_{Bare}^2 \!+\!\lambda^2 \HVEV^2 
\!-\!  C^{Unrenorm;1-loop;\Lambda^2}_{\LSMmth}\Lambda^2  ]~ \vec \pi \nonumber
\\ &&=- \HVEV \mpisq \vec \pi \nonumber
\\ &&\xrightarrow[\HVEV \neq 0, \mpisq\rightarrow0]{}  - \HVEV m^{2}_{\pi ;NGB;\GML} \vec \pi \nonumber
\\ &&\equiv 0
\end{eqnarray}
{\em But, because O(4) is spontaneously broken, a new NGB $SU(2)_{L-R}$ chiral symmetry  arises in the Goldstone mode  limit, so the conservation of the axial-vector current in  (\ref{PCAC1-loopUVQDGoldstone})  cannot be considered to be  fine-tuned.}

{\bf Restoration of Nambu Goldstone boson $SU(2)_{L-R}$ chiral symmetry in the spontaneously broken limit of the  \GMLfull \LSM:} Following \cite{LynnStarkman2013a}, we re-write (\ref{eqn:Eff-1loop-GM-orig})  in the unitary Kibble representation of  $\Phi$ \cite{Ramond2004,Georgi2009}:
\begin{eqnarray}
\label{eqn:LGoldstone1-loopUnitary}
L^{Effective;1-loop;\Lambda^2}_{Goldstone;\LSMmth} &=& -\frac{1}{2} \left( \partial_{\mu} {\tilde H} \right)^2 - \frac{1}{4} {\tilde H}^2 Tr \left[ \partial_{\mu} U^{\dagger} \partial_{\mu} U \right] \nonumber
\\ && -\frac{1}{4} \lambda^2\left[ {\tilde H}^2 - \HVEV^2 \right]^2 \,,
\end{eqnarray} 
which restores the inhomogeneous NGB  $SU(2)_{L-R}$ chiral symmetry 
${\vec{\tilde\pi}} \rightarrow {\vec{\tilde\pi}} + {\HVEV {\vec {\theta}}} ~+ {\cal O}(\theta^2)$, thus restoring the full $SU(2)_L \times SU(2)_R \sim O(4)$ symmetry. 
The ${\vec{\tilde\pi}}$ are true NGB in (\ref{eqn:LGoldstone1-loopUnitary}) with only derivative couplings.
{\em By definition, the restoration of a symmetry cannot be considered fine-tuning, so neither can the Goldstone mode results of this paper!}

{\bf Higgs mass-squared $m_h^2$ is not fine-tuned in the Goldstone mode of  the \GMLfull \LSM:} A central observation of this paper is that,  in consequence of SSB, 
careful definition of the 1-loop-UVQD-improved vacuum, 
automatic (Ward-Takahashi identity enforced)  Higgs-VSC tadpole renormalization, 
and Goldstone-SRC enforcement of the Goldstone Theorem, 
{\em all finite remnants of 1-loop UVQD contributions to $m_h^2$ are absorbed into the dimension-2 relevant operator
$\frac{1}{2}m^2_{\pi;NGB;\LSMmth} h^2$
in (\ref{eqn:V-Renorm-1loop-Lambda2-GM}).
Therefore all 1-loop UVQD  contributions to $m_h^2$ vanish identically with $m^2_{\pi;NGB;\LSMmth} \equiv 0$ in (\ref{eqn:V-Renorm-1loop-Lambda2-GM})}.  $L^{Effective;1-loop;\Lambda^2}_{Goldstone;\LSMmth}$ 
  then gives the sensible Higgs mass
\be
\label{eqn:NoHiggs-FTP1Loop}
  m_h^2 = 2\lambda^2 \HVEV^2 .
 \ee
$m_h^2$ is here at worst logarithmically divergent, because $\HVEV$ receives no UVQD \cite{Lee1970,Symanzik1970a,Symanzik1970b,Vassiliev1970,ItzyksonZuber1980}. But (crucially) we have traced equation (\ref{eqn:NoHiggs-FTP1Loop}) to its origin, the restoration of inhomogeneous NGB  $SU(2)_{L-R}$ chiral symmetry ${\vec{\tilde\pi}} \rightarrow {\vec{\tilde\pi}} + {\HVEV {\vec \theta}}+{\cal O} (\theta^2)$ in the Goldstone mode of the \GML \LSM. {\em Because it is the result of the exact NGB  $SU(2)_{L-R}$ chiral symmetry, the Higgs mass-squared in (\ref{eqn:NoHiggs-FTP1Loop}) cannot be considered fine-tuned! The \GMLfull \LSM~ therefore has no Higgs Fine Tuning Problem in its Goldstone mode.}

Before going forward, we now summarize our 1-loop results.
We emphasize that most of the calculations in Section \ref{sec:O4LSM} (with the exception, for example, of the inclusion of 1-loop UVQD from virtual Standard Model quarks and leptons in subsection \ref{sbscn:withSMQL-noHiggs-FTP}) are not new;
 rather, they have been common  knowledge for more than four decades. 
 What is new are {\bf certain observations that apply directly  to the proper 1-loop UVQD renormalization of the spontaneously broken O(4) \LSM}~  (i.e. \GMLfull model in Goldstone mode):
 \begin{enumerate}
 \item The Higgs VEV $\HVEV$ receives no UVQD, only logarithmic UV divergences from wave-function renormalization \cite{Lee1970,Symanzik1970a,Symanzik1970b,Vassiliev1970,ItzyksonZuber1980}.
 \item {\bf Proper renormalization of the \GMLfull \LSM:} Because the origin is singular in $\mu ^2$, B.W. Lee \cite{Lee1970}, K. Symanzik \cite{Symanzik1970a,Symanzik1970b}, A. Vasiliev \cite{Vassiliev1970}, C. Itzykson and J-B. Zuber \cite{ItzyksonZuber1980} and \cite{LynnStarkman2013a} argue strongly that
one {\bf must} implement the renormalization of the theory throughout the entire $\vert\HVEV\vert$ vs. $\mpisq/\lambda^2$ quarter-plane in Figure \ref{fig6}: this is a requirement, not a choice, for proving the renormalizabilty of the spontaneously broken O(4)\LSM. 
\item Spontaneously broken O(4) \LSM~  must therefore be viewed as a limiting case of the  \GMLfull model's generic set of  $\mpisq/\lambda^2\geq0$ theories 
(i.e. the  $\mpisq/\lambda^2=0$  line (y-axis) in the  $\vert\HVEV\vert$ vs. $\mpisq/\lambda^2$  quarter-plane shown in Figure \ref{fig6}), where Lee and Symanzik's Goldstone-SRC (a renormalization prescription) must be explicitly imposed on the O(4) \LSM~ Goldstone-mode vacuum and excited states:
	\begin{quote}
	{\bf Goldstone Symmetry Restoration Condition:} the masses of Nambu-Goldstone Bosons must be fixed to exactly zero in the Goldstone-mode (y-axis) to enforce the Goldstone Theorem. 
	\end{quote}
\item An additional Lee/Symanzik renormalization condition  on the O(4) \LSM~vacuum, Higgs-VSC, is automatically enforced by Ward-Takahashi identities. Otherwise it would have been necessary to impose it explicitly as in \cite{Lee1970,ItzyksonZuber1980}:
	\begin{quote}
	{\bf Higgs Vacuum Stability Condition:} everywhere in the $\vert\HVEV\vert-\mpisq/\lambda^2$ quarter-plane the Higgs must not simply disappear into the vacuum;
	\end{quote}
\item Ward-Takahashi identities (including the automatic vanishing of  tadpoles)  force all UVQD in the S-Matrix to be absorbed 
	into the physical renormalized pseudo-scalar pion (pole) mass-squared $\mpisq$ everywhere in the quarter-plane.
\item All relevant dimension-2 and dimension-1 operators (with UVQD coefficients) form a renormalized Higgs potential  (Figure \ref{fig5})
	that is well-defined everywhere in the quarter-plane and is minimized at the UVQD-corrected vacuum $H=\HVEV ,\vec \pi =0$.  
\item	Although the coefficients of the dimension-2 scalar self-energy 2-point, and dimension-1 tadpole-1-point, relevant operators 
	take their ÒnaturalÓ scale   $\mu_{Bare}^2 \sim \Lambda^2$,
	there is still no need to fine-tune away the UVQD to get a Òweak-scale HiggsÓ mass-squared $m_h^2G^{Exp}_\mu\simeq O(1)$  if experiment so demands, 
	once $\mpisq\to m^2_{\pi;NGB;\LSMmth}\equiv0$, is fixed.
\item The Goldstone Theorem insists that the NGBs are {\em exactly} massless.  
          Since all UVQD are absorbed into NGB masses-squared $m^2_{\pi;NGB;\LSMmth}$, they vanish identically and exactly as  $m^2_{\pi;NGB;\LSMmth} \to 0$.
\item	This forces {\em all} UVQD in the S-Matrix to vanish exactly, with finite remnant exactly zero. They are not absorbed into renormalized parameters,
	including, but especially, $m_h^2$.
\item The root cause of this fine-tuning-free disappearance of UVQD is  the imposition of the Lee/Symanzik Goldstone-SRC on the vacuum and excited states of the spontaneously broken theory.  
\item Our no-fine-tuning-theorem for a weak-scale Higgs mass (in the spontaneously broken O(4) \LSM) 
	is then simply another (albeit un-familiar) consequence of the Goldstone Theorem, 
	an exact property of the spontaneously broken  vacuum and spectrum.
\item Along any path in the quarter-plane of Figure \ref{fig6}, either smooth (like $\alpha \delta$) or not (like $\alpha \beta \gamma \delta$), from the Wigner mode to the Goldstone mode, Ward-Takahashi identities automatically enforce (including all UVQD) the Higgs-VSC, and the physical Higgs particle is prevented from simply disappearing into the vacuum. The vacuum is always stable (i.e. the theory is stationary) at the minimum of the effective potential (\ref{eq:form}) at $H=\HVEV$ and $\vec \pi =0$.

\item ``New" UVQD corrections to the PCAC relation (widely unfamiliar to modern audiences) arise in (\ref{PCAC1-loopUVQDGoldstone}) in the \GMLfull ($\epsilon \neq 0$) model, but not in the Schwinger ($\epsilon \equiv 0$) model \cite{Schwinger1957}. These are guaranteed (by Ward-Takahashi identities) to be exactly the same as the UVQD corrections to the renormalized potential (\ref{eq:form1loop}). Therefore they are also absorbed into $\mpisq$
and  vanish (without fine-tuning) as $m_{\pi ;NGB;\GML}^2 \to 0$ in Goldstone mode.

\item Inhomogeneous Nambu Goldstone boson $SU(2)_{L-R}$ symmetry (i.e. where, to lowest order,  NGB are shifted by constants ${\vec{\tilde\pi}} \rightarrow {\vec{\tilde\pi}} + {\HVEV {\vec \theta}}+{\cal O}(\theta^2)$) is explicitly broken by the \GMLfull PCAC relation in Figure \ref{fig6}, everywhere but the y-axis. On the y-axis (i.e. in the spontaneously broken theory) this $SU(2)_{L-R}$ chiral symmetry is restored. The restoration of an exact symmetry is the strongest definition of ``no fine-tuning".

\item Axial-vector current conservation $\partial_\mu {\vec J}^5_\mu\to0$ is restored in the spontaneously broken limit  of the \GMLfull \LSM~ (the y-axis of Figure \ref{fig6}), a direct consequence of the restoration of NGB $SU(2)_{L-R}$ chiral symmetry there.  By definition, current conservation cannot be considered fine-tuning. Because we have traced them to {\em exact} NGB $SU(2)_{L-R}$ chiral symmetry, neither can the other results of this paper on the y-axis of Figure \ref{fig6}. For example, all UVQD vanish exactly in the spontaneously broken limit of   \GML \LSM~without fine-tuning.

\item All finite remnants of 1-loop UVQD contributions to $m_h^2$ are absorbed into $m^2_{\pi;NGB;\LSMmth}$ and vanish identically as $m^2_{\pi;NGB;\LSMmth} \equiv 0$.
{\em But, because it is the result of the exact NGB $SU(2)_{L-R}$ chiral symmetry, the Higgs mass-squared in (\ref{eqn:NoHiggs-FTP1Loop}) cannot be considered fine-tuned. The \GMLfull \LSM~ therefore has no Higgs Fine Tuning Problem in its Goldstone mode.}

\item An important subtlety, the so-called ``Higgs fine-tuning discontinuity" \cite{LynnStarkman2013a} arises, here in O(4), when one compares the UVQD renormalization of the Schwinger ($\epsilon \equiv 0$) model \cite{Schwinger1957}, to the spontaneously broken limit of the UVQD renormalization of the \GMLfull ($\epsilon \neq 0$) model.  We have argued in this paper (in agreement with B.W. Lee, K. Symanzik, A. Vassiliev, C. Itzykson and J-B. Zuber
 \cite{Lee1970,Symanzik1970a,Symanzik1970b,Vassiliev1970,ItzyksonZuber1980} and  \cite{LynnStarkman2013a}), that the {\em correct} procedure is to: {\em first} renormalize the \GML O(4)\LSM~everywhere in  the $\vert\HVEV\vert$ vs. $\mpisq /\lambda^2$ quarter-plane of Figure \ref{fig6}; {\em then} take the spontaneously broken (y-axis) limit. {\em A comparative discontinuity in the Higgs mass arises because, at the UVQD level, the Schwinger model is not the same as the $\mpisq \to 0$ limit of the \GMLfull model.} (This is reminiscent \cite{LynnStarkman2013a} of other such discontinuities: explicit photon masses \cite{vanDamVeltman1970}; explicit graviton masses \cite{vanDamVeltman1970,Zakharov1970}; in gauge theories of massive spin-1 bosons, an explicit mass term generates a very different UVQD structure and high energy behavior than does the Higgs mechanism \cite{JCTaylor1976}.)

 \item Resolution of the UVQD Higgs fine-tuning discontinuity motivates and defines the {\bf Goldstone mode renormalization prescription  (GMRP) \cite{LynnStarkman2013a}:} when calculating with the Schwinger model bare Lagrangian \cite{Schwinger1957}, one must impose (to each loop order, essentially by hand) B.W. Lee and K. Symanzik's two renormalization conditions \cite{Lee1970,Symanzik1970a,Symanzik1970b}:
 \begin{quote}
	{\bf Higgs-VSC:}  The physical Higgs particle must not simply disappear into the vacuum.
	\end{quote}
 \begin{quote}
	{\bf Goldstone-SRC: } Nambu-Goldstone Bosons masses must be fixed to exactly zero 	\end{quote}
The GMRP is particularly useful in the renormalization of the Standard Model \cite{JCTaylor1976,Lynn2011}.

\item There is no possibility (or need) to cancel UVQD between virtual bosons and fermions in the  O(4) \LSM.  (After all, we have so far not allowed for any fermions, but will introduce later, in subsection \ref{sbscn:withSMQL-noHiggs-FTP} and Appendix C.)
\item It is un-necessary to impose any new ``Beyond the O(4) \LSM " symmetries or other ``new physics": 
	weak-scale spontaneously broken O(4) \LSM~   already has sufficient symmetry to force all S-Matrix UVQD to vanish and to ensure that it does not suffer a Higgs-FTP.
\end{enumerate}

The 1-loop results enumerated above (as well as various others) will be generalized in the remainder of this paper as follows:
\begin{itemize}
\item  To all-loop orders of perturbation theory for  the \GMLfull \LSM~of scalars, in subsection \ref{sbscn:all-loop-noHiggs-FTP} and Appendix B.
\item  To 1-loop for  the $SU(2)_L$  \LSM~ with PCAC of scalars coupled to Standard Model quarks and leptons, in subsection \ref{sbscn:withSMQL-noHiggs-FTP} and Appendix C.
\item  To all-loop orders of perturbation theory for  the $SU(2)_L$  \LSM~with PCAC of scalars coupled to Standard Model quarks and leptons, in subsection \ref{sbscn:withSMQL-noHiggs-FTP} and Appendix C.
\end{itemize}

We would also remind the reader of results specific to 1-loop: 
\begin{itemize}
\item Following M.J.G. Veltman \cite{Veltman1981}, Appendix D shows that  our 1-loop results do not depend on choice of UV regularization scheme (e.g. n-dimensional or  UV cut-off).
\item  The fact that all 1-loop UVQD in the SM Higgs self-energy and mass vanish exactly after tadpole renormalization has been known \cite{Lynn1982b,Stuart1985}
	for more than 3 decades.
\end{itemize}

\subsection{Goldstone Exception: Weak-scale spontaneously broken O(4) \LSM~  has, to all orders of loop perturbation theory, no Higgs-FTP}
\label{sbscn:all-loop-noHiggs-FTP}

The reader should worry that the vanishing of 1-loop UVQD is insufficient to demonstrate that the theory does not require fine-tuning. 
UVQD certainly appear at multi-loop orders and fine-tuning  $\mu_{Bare}^2$ might yet be required. 
If each loop order contributed a factor $\hbar/16\pi^2\simeq10^{-2}$,  
then cancellation of UVQD to more than 16 1PI loops would be required to defeat a factor of  $G^{Exp}_\mu \Lambda^2\leq10^{-32}$ for $\Lambda\geq M_{Pl}$. 

B.W. Lee, and K. Symanzik  renormalized generic O(4) \LSM~ (i.e. in the full $\vert\HVEV\vert-\mpisq/\lambda^2$ quarter-plane of Figure \ref{fig6}) 
to all-loop-orders more than forty years ago. We simply remind the reader of various results \cite{Lee1970,Symanzik1970a,Symanzik1970b,Vassiliev1970,Symanzik1969,Gervais1969,Lee1969,Bogoliubov1957,Hepp1966,ZimmermanDeser} relevant to this paper. For a complete discussion in modern functional integral language, see the textbook by C. Itzykson and J-B. Zuber \cite{ItzyksonZuber1980}, chapter 11-4. The UV divergence and counter-term structure must be the same throughout the  $\vert\HVEV\vert$ vs. $\mpisq/\lambda^2$  quarter-plane (Figure \ref{fig6}), 
 	giving proper interpolation between the symmetric Wigner mode ($\HVEV=0$, $\mpisq\neq0$) 
	and the spontaneously broken Goldstone mode ($\HVEV\neq0$, $\mpisq=0$). Appendix B gives details of the renormalization of the O(4)\LSM~ by UVQD to all orders of loop perturbation theory.

{\bf Ward-Takahshi identity:} B.W. Lee,  K. Symanzik proved \cite{Lee1970,Symanzik1970a,Symanzik1970b,Vassiliev1970} that the Ward-Takahashi identity (\ref{eqn:WTidentity}) (which governs the 2-pion connnected Greens function and sets the strength of the PCAC 
relation) and the PCAC relation itself (\ref{eqn:DivergenceofAmu}) receive no UVQD corrections to all-loop-orders of perturbation theory.
Using the O(4) Ward-Takahashi identity  \cite{Lee1970,Symanzik1970a,Symanzik1970b,Vassiliev1970,ItzyksonZuber1980}, the UVQD corrections to (\ref{eqn:LPCAC}) can   be included in the renormalized explicit symmetry breaking term
\begin{equation}
\label{eqn:LSymmetryBreaking_All-loops}
L^{All-loops; PCAC}_{\LSMmth} = \epsilon^{All-loops}_{\LSMmth} H\,,
\end{equation}
\be
\label{eqn:EpsilonAll-Loops}
\epsilon^{All-loops}_{\LSMmth}  = \HVEV\mpisq
\ee
Here 
$\mpisq$ is the physical renormalized pion (pole) mass for three degenerate pions,
and includes {\em exactly the all-loop-orders UVQD contributions from virtual scalars  in  (\ref{eqn:LAllLoopsLambdaSquared} and \ref{eqn:MpiAllLoop}) below}, while neglecting logarithmic divergences and other finite contributions.

{\bf Higgs Vacuum Stability Condition:}
Following \cite{LynnStarkman2013a}, the O(4) Higgs-VSC is then automatically enforced by Ward-Takahashi identities (in Appendix B),
so $\langle h \rangle =0$ everywhere in the $\vert\HVEV\vert-\mpisq/\lambda^2$ quarter-plane and the physical Higgs particle cannot simply disappear into the vacuum.

{\bf Renormalized Potential:}  Appendix B shows that Ward-Takahashi identity-enforced Higgs-VSC (i.e. all-orders tadpole renormalization) is sufficient to force all S-matrix UVQD into the renormalized pion pole mass  $\mpisq$ in the all-loop-orders UVQD-corrected renormalized potential in (\ref{eqn:VAllLoop})
 \begin{eqnarray}
 \label{eqn:VAllLoopIsVTree}
V^{Renorm;All-loops;\Lambda^2}_{\LSMmth} &=& \lambda^2 \left[ \Phi^\dagger\Phi - \half\left( \HVEV^2 - \frac{\mpisq}{\lambda^2}\right)\right]^2\nonumber \\
& &- \HVEV \mpisq h
\\ &=& V^{Tree}_{\LSMmth} \,,
\end{eqnarray}
for fixed $\mpisq$.
A cross-section of $V^{Renorm;All-loops;\Lambda^2}_{\LSMmth} $, through its absolute minimum (the vacuum $h=\HVEV, \langle \vec \pi \rangle=0$) and its local maximum, is plotted in Figure \ref{fig5}. It is identical to the tree-level result (\ref{eq:form}).

{\bf Goldstone-Symmetry Restoration Condition:} 
NGB masses must be identically and exactly zero:  \begin{flalign}
\quad\quad&\mpisq \equiv  \mu_{Bare}^2  +\lambda^2\HVEV^2 &\notag\\
\quad\quad&\quad- C^{Unrenorm;All-loops;\Lambda^2}_{\LSMmth}(\HVEV^2,\lambda^2,\mpisq,m_h^2) \Lambda^2& \notag\\
\quad\quad&\quad \xrightarrow[\HVEV \neq 0, \mpisq\rightarrow0]{}
\\ \quad\quad&m^2_{\pi;NGB;\LSMmth} \equiv \mu_{Bare}^2 \nonumber +\lambda^2\HVEV^2\notag\\
\quad\quad&\quad - C^{Unrenorm;All-loops;\Lambda^2}_{\LSMmth}(\HVEV^2,\lambda^2,0,2\lambda^2\HVEV^2) \Lambda^2 \notag\\
\quad\quad&\quad\equiv 0\notag
\end{flalign}
Here 
\begin{eqnarray}
\label{LAllLoopsLambdaSquare}
&& L^{All-loops;\Lambda^2}_{\LSMmth} = \nonumber
\\ && C^{\cdots}_{\LSMmth} \left(\HVEV^2,\lambda^2,\mpisq,m_h^2=\mpisq+2\lambda^2\HVEV^2\right) \Lambda^2 \nonumber
\\ && \times \left[\frac{h^2}{2}  + \frac{\pi_3^2}{2} + \pi_+\pi_- + \HVEV h \right]
\end{eqnarray}
 is the all-loop UVQD Lagrangian (\ref{eqn:LAllLoopsLambdaSquared}) arising from Feynman diagrams, the analogy of the 1-loop result (\ref{eqn:L-1loop-Lsq-LSM}).
$C^{Unrenorm;All-loops;\Lambda^2}_{\LSMmth}$ is  finite and constant (because of the dimensionality of  (\ref{LAllLoopsLambdaSquare}))
and dependent (because of nested divergences within multi-loop 1PI graphs) on the finite constant physical input parameters of the theory. 

Therefore
\begin{eqnarray}
\label{eqn:VrenormAllloopGM}
 V^{Renorm;All-loops;\Lambda^2}_{Goldstone;\LSMmth}&& = \lambda^2\left[\frac{h^2}{2}  + \frac{\pi_3^2}{2} + \pi_+\pi_- + \HVEV h \right]^2  \nonumber \\
&&= \lambda^2\left[\Phi^\dagger\Phi^2 - \frac{\HVEV^2}{2} \right]^2\,. 
  \end{eqnarray}

{\bf Restoration of Nambu Goldstone boson $SU(2)_{L-R}$ chiral symmetry in the spontaneously broken limit of the  \GMLfull \LSM:} Now re-write (\ref{eqn:VrenormAllloopGM})  in the unitary Kibble representation of  $\Phi$ \cite{Ramond2004,Georgi2009} and form the all-loop-orders effective Lagrangian:
\begin{eqnarray}
\label{eqn:LGoldstoneAll-loopUnitary}
L^{Effective;All-loop;\Lambda^2}_{Goldstone;\LSMmth} &=& -\frac{1}{2} \left( \partial_{\mu} {\tilde H} \right)^2 - \frac{1}{4} {\tilde H}^2 Tr \left[ \partial_{\mu} U^{\dagger} \partial_{\mu} U \right] \nonumber
\\ && -\frac{1}{4} \lambda^2\left[ {\tilde H}^2 - \HVEV^2 \right]^2 
\\ &=& L^{Tree}_{Goldstone;\LSMmth} \,,\nonumber
\end{eqnarray} 
which of course has the inhomogeneous NGB $SU(2)_{L-R}$ chiral symmetry 
${\vec{\tilde\pi}} \rightarrow {\vec{\tilde\pi}} + {\HVEV {\vec \theta}}~+ {\cal O}(\theta^2)$. The ${\vec{\tilde\pi}}$ are true NGB with only derivative couplings.
{\em By definition, the restoration of an exact symmetry cannot be considered fine-tuning, so neither can the all-loop-orders Goldstone-mode results in this paper.}

{\bf UVQD corrections to the \GMLfull PCAC relation in Goldstone mode:} Following \cite{LynnStarkman2013a}, we show that the ``new" (i.e. widely unfamiliar to modern audiences) all-loop-orders UVQD corrections to the O(4) PCAC relation also vanish (without fine-tuning) in Goldstone mode:
\begin{eqnarray}
\label{PCACAll-loopUVQDGoldstone}
\partial_{\mu} {\vec J}^{5}_{\mu} &&= -\HVEV [ \mu_{Bare}^2 \!+\!\lambda^2 \HVEV^2 
\!-\!   C^{Unrenorm;All-loop;\Lambda^2}_{\LSMmth}\Lambda^2  ]~ \vec \pi \nonumber
\\ &&= - \HVEV \mpisq \vec \pi \nonumber
\\ &&\xrightarrow[\HVEV \neq 0, \mpisq\rightarrow0]{}  - \HVEV m^{2}_{\pi ;NGB;\GML} \vec \pi 
\\ &&\equiv 0 \nonumber
\end{eqnarray}
Because  $SU(2)_{L-R}$ chiral symmetry  is exact (to all-loop-orders UVQD) in the Goldstone mode  limit of the \GMLfull \LSM, the conservation of the axial-vector current in  (\ref{PCACAll-loopUVQDGoldstone})  cannot (by definition) be considered to be  fine-tuned.

{\bf Higgs mass-squared $m_h^2$ is not fine-tuned in the all-UVQD-loop-orders Goldstone mode of  the \GMLfull \LSM:} 
With Goldstone SRC,  all UVQD contributions to $m_h^2$ are absorbed into the dimension-2 relevant operator
$\frac{1}{2}m^2_{\pi;NGB;\LSMmth} h^2$
in  (\ref{eqn:VAllLoopIsVTree}) and vanish identically with $m^2_{\pi;NGB;\LSMmth} \equiv 0$. $V^{Renorm;All-loops;\Lambda^2}_{Goldstone;\LSMmth}$ in (\ref{eqn:VrenormAllloopGM}) 
  then gives the sensible Higgs mass
\be
\label{NoHiggs-FTP}
  m_h^2 = 2\lambda^2 \HVEV^2 \,,
 \ee
which (because $\HVEV$ receives no UVQD in any order of perturbation theory \cite{Lee1970,Symanzik1970a,Symanzik1970b,Vassiliev1970,ItzyksonZuber1980}) is at worst logarithmically divergent. We have therefore traced (\ref{NoHiggs-FTP}) to the restoration of inhomogeneous $SU(2)_{L-R}$ chiral symmetry  in the Goldstone mode of the \GML \LSM. {\em Because it is the result of an exact symmetry, the Higgs mass-squared in (\ref{NoHiggs-FTP}) cannot be considered fine-tuned!} The \GMLfull \LSM~ therefore has no Higgs Fine Tuning Problem in its Goldstone mode to all orders in loop perturbation theory

The results enumerated at the end of subsection \ref{sbscn:1loop-noHiggs-FTP} therefore generalize to all-loop-orders.

 \subsection{ $SU(2)_L$ \LSM~with PCAC in the   $\vert\HVEV\vert$ vs. $\mpisq/\lambda^2$  quarter-plane with SM quarks and leptons }
 \label{sbscn:withSMQL-noHiggs-FTP}

All-orders renormalization of the generic bosonic O(4) \LSM~ was extended to include parity-doublet nucleons by J.L Gervais  and  B.W.Lee \cite{Gervais1969}. 
The full reasoning necessary to secure renormalized inclusion of SM quarks and leptons in the $\vert\HVEV\vert$ vs. $\mpisq/\lambda^2$  quarter-plane 
is beyond the scope of this paper (e.g. involves infra-red subtleties \cite{Lee1970,Symanzik1970a,Symanzik1970b,Vassiliev1970,Symanzik1969,Gervais1969,Lee1969,Bogoliubov1957,Hepp1966,ZimmermanDeser,Bjorken1965,Peskin1995,Kaku1993}).
We merely remind the reader of the key complication, that SM fermion masses vanish in the symmetric Wigner mode ($\HVEV=0$, $\mpisq/\lambda^2\neq0$), as
\be
m_{quark},m_{lepton} \propto \HVEV \,.
\ee
If we couple SM quarks and leptons (including Dirac-mass neutrinos) to the $SU(2)_L$ \LSM~ in the usual way, 
then new UVQD (from virtual fermions) are proportional to Yukawa couplings squared $y_{i}^2$, and 
appear only in  coefficients of relevant dimension-2 and dimension-1 scalar-sector operators and in the renormalized PCAC relation. 
They are assembled and included in the scalar-sector effective   Lagrangian in Appendix C (equation (\ref{eqn:fermioncontribution})).

{\bf Feyman diagrams} We focus on explicit  calculation of 1-loop UVQD contributions, due to SM quarks and leptons,
in the generic   $\vert\HVEV\vert$ vs. $\mpisq/\lambda^2$  quarter-plane. 

The total UVQD 1-loop  Lagrangian is
\begin{eqnarray}	
\label{eqn:fermioncontributionBodyPaper}
L^{1-loop;\Lambda^2}_{\LSMmth+SM q\& l} &=& 
\\ &&\!\!\!\!\!\!\!\!\!\!\!\!\!\!\!\!\!\!\!\!\!\!\!\!\!\!\!\!\!\!
C^{Unrenorm;1-loop;\Lambda^2}_{\LSMmth+SM q\& l}  \Lambda^2 \left(\Phi^\dagger\Phi - \half\HVEV^2\right)\nonumber
\end{eqnarray}
With quark and lepton masses-squared $m_i^2 = \half y_i^2 \HVEV^2$, the coefficient in (\ref {eqn:fermioncontributionBodyPaper}) (calculated in Appendix C, equation (\ref{eqn:app-to-quote-2}))
 can be written \cite{Veltman1981}
\begin{eqnarray}
\label{eqn:C1LoopLSMSMQandL}
 C^{Unrenorm;1-loop;\Lambda^2}_{\LSMmth+SM q\& l} &\\
 & \!\!\!\!\!\!\!\!\!\!\!\!\!\!\!\!\!\!\!\!\!\!\!\!\!\!\!\!\!\!\!\!\!\!\!
 = \left(  \frac{  -3 m_h^2 + 4\mathop{\sum}_{quarks}m_q^2  + 4\mathop{\sum}_{leptons}m_l^2 }{16\pi^2\HVEV^2}\right) \nonumber
\end{eqnarray}
where the sums are over flavor and, for the quarks, color.
Equation (\ref{eqn:fermioncontributionBodyPaper}) is famously recognizable as the zero-gauge-coupling limit of 1-loop UVQD in the SM \cite{Veltman1981,Lynn1982b,Stuart1985},
 where to this order $-6\lambda^2 \HVEV^2=-3m_h^2$.

{\bf Symmetry breaking $O(4) \to SU(2)_L$:} Standard Model Yukawa couplings reduce the symmetry of the theory because of the mass-splittings within left-handed quark and lepton doublets.

{\bf Partially conserved axial-vector currents:} Apart from  an explicit symmetry breaking term
 (\ref{eqn:LPCACSMQandL}) 
 below, 
the theory has $SU(2)_L$  symmetry after inclusion of Standard Model fermions: left-handed transforming as doublets and right-handed transforming as singlets. Since $SU(2)_L$ has 3 generators, there are only 3 currents: the CVC  
and PCAC together  form a single $SU(2)_L$  {\em isospin} current, where the inclusion of SM quarks and leptons modifies the scalar $SU(2)_L$ \LSM~ current with the usual left-handed fermion bi-linears:
\begin{eqnarray}
\label{eqn:IsospinCurrent}
\vec{J}_\mu^{Isospin} &=& \vec{J}_\mu^{Scalars} + \vec{J}_\mu^{Fermions}\,, \nonumber \\ 
\vec{J}_\mu^{Scalars} &=& \vec{\pi} \times \partial_\mu \vec{\pi} + \vec{\pi}\partial_\mu H - H \partial_\mu\vec{\pi}\,,\nonumber \\ 
\vec{J}_\mu^{Fermions} &=& \vec{J}_\mu^{Quarks}+\vec{J}_\mu^{Leptons}\,  \\ 
\vec{J}_\mu^{Quarks} &=& \mathop{\sum}_{flavor,color} \! \bar q\gamma_{\mu} \vec \sigma \left(\frac{1+\gamma^5}{2}\right) q\,,
\nonumber 
\\  \vec{J}_\mu^{Leptons} &=& \mathop{\sum}_{flavor}\! \bar l\gamma_{\mu} \vec \sigma \left(\frac{1+\gamma^5}{2}\right) l \,. \nonumber
\end{eqnarray}

With the explicit symmetry breaking term
\be
\label{eqn:LPCACSMQandL}
L^{1-loop;PCAC}_{\LSMmth+SM q \& l} = \epsilon_{\LSMmth+SM q \& l}^{1-loop}~H
\ee
the divergence of the isospin current becomes \cite{Lynn2011}
\begin{eqnarray}
\label{eqn:SBCurrentDivergencesIsospin}
\partial_\mu\vec{J}_\mu^{Isospin} &=& -\epsilon_{\LSMmth+SM q\& l}^{1-loop}~  {\vec \pi}
\end{eqnarray}

{\bf Higgs VEV:} B.W. Lee and K. Symanzik \cite{Lee1970,Symanzik1970a,Symanzik1970b,Vassiliev1970} proved, in the scalar \GMLfull model of strongly interacting pions and a scalar sigma particle, that the Higgs VEV $\HVEV$
receives no UVQD to all orders in perturbation theory. Lee's proof is outlined in \cite{LynnStarkman2013a}. B.W. Lee and J-L. Gervais \cite{Gervais1969} extended that proof to include protons and neutrons. 
Using naive power counting of the degree of UV divergence of the coefficients of 1PI operators, \cite{LynnStarkman2013a} proves that $\HVEV$ receives no UVQD, in its toy U(1) \LSM~ with PCAC, including the effects of a Dirac fermion with Yukawa coupling. That proof is easily extended to include Standard Model quarks and lepton \cite{Lynn2011}. 

{\bf Ward-Takahashi identity:} In its toy U(1) with PCAC \LSM~ with a single fermion, \cite{LynnStarkman2013a} proves that the disposition of all 1-loop UVQD corrections, to the dimension-1 relevant symmetry breaking operator (i.e. the U(1) version of (\ref{eqn:LPCACSMQandL})),
is controlled by the Ward-Takahashi identity \cite{LynnStarkman2013a} governing the pion connected 2-point Greens function, and that this 
determines (for UVQD purposes) the strength of the renormalized  symmetry breaking term. 

Those results are extended to $SU(2)_L$ with SM fermions in \cite{Lynn2011}. The UVQD corrections inside (\ref{eqn:LPCACSMQandL}) then give the renormalized explicit symmetry breaking term
\be
\epsilon^{1-loop}_{\LSMmth+SM q \& l}  = \HVEV\mpisq
\ee
Here 
$\mpisq$ is the physical renormalized pion (pole) mass for three degenerate pions,
and includes {\em exactly the 1-loop UVQD contributions from virtual scalars, quarks and leptons in (\ref{eqn:C1LoopLSMSMQandL})} \cite{Lynn2011} (neglecting logarithmic divergences and other finite contributions).

{\bf Higgs Vacuum Stability Condition: } Then, following \cite{LynnStarkman2013a} but extended here to $SU(2)_L$, the Higgs-VSC is automatically enforced by Ward-Takahashi identities,
so that the Higgs $\langle h \rangle =0$ everywhere in the $\vert\HVEV\vert-\mpisq/\lambda^2$ quarter-plane and the physical Higgs particle cannot simply disappear into the vacuum, as demonstrated in (\ref{V1LoopLSMSMQandL}).

{\bf Renormalized potential:} Appendix C (equation (\ref{V1LoopLSMSMQandL})) shows that, after Ward-Takahashi-identity-enforced Higgs-VSC tadpole renormalization, 
 the renormalized potential, 
 keeping 1-loop UVQD (but ignoring un-interesting logarithmic divergences, finite parts and vacuum energy/bubbles) 
 can again be written (in analogy with subsection \ref{sbscn:1loopUVQD}):
\begin{eqnarray}
\label{eqn:1loopEffectiveLwithQL}
 V^{Renorm;1-loop;\Lambda^2}_{\LSMmth +SM q\& l}  &=&
\\  \lambda^2 \Bigl[\Phi^\dagger\Phi &-& \half\left(\HVEV^2 - \frac{\mpisq}{\lambda^2}\right)\Bigr]^2 - \HVEV \mpisq H \nonumber
\\ &=& V^{Tree}_{\LSMmth}\,,
\end{eqnarray}
for fixed $\mpisq$,
with $H = h + \HVEV, \langle h\rangle=0$.
Here
\be
\label{MPiUVQDSMQandL}
\mpisq \! =  \mu_{Bare}^2 \!-\! C^{Unrenorm;1-loop;\Lambda^2}_{\LSMmth+SM q\& l}\Lambda^2 + \lambda^2\HVEV^2
\ee
is, everywhere in (\ref{eqn:1loopEffectiveLwithQL}), {\em still} the 
physical renormalized pseudo-scalar pion (pole) mass-squared,
and 
\be
\label{eqn:mhsq_SMql_1loop}
m_h^2 = \mpisq + 2\lambda^2 \HVEV^2 \geq \mpisq.
\ee
Once again, via (\ref{eqn:1loopEffectiveLwithQL}), (\ref{MPiUVQDSMQandL}) and (\ref{eqn:mhsq_SMql_1loop}),
 the physical renormalized pseudo-scalar pion pole mass-squared $\mpisq$  has absorbed all 1-loop  UVQD, 
 which now include those from virtual SM quarks and leptons. 
 
{\bf 1-loop Nambu Goldstone bosons}:
	It is obvious that $V^{Renorm;1-loop;\Lambda^2}_{\LSMmth+SM q\& l} $  in (\ref{eqn:1loopEffectiveLwithQL})  and Figure~2
	has Nambu-Goldstone Bosons only when ``bottom-of-the-wine-bottle Goldstone symmetry" is restored, i.e. in Goldstone mode on the  $\mpisq=0$ line (y-axis of Figure \ref{fig6}). 
	After Goldstone-SRC, the Goldstone Theorem forces all 1-loop UVQD in (\ref{eqn:C1LoopLSMSMQandL}) and their finite remnants to vanish exactly and identically: 
	\begin{eqnarray}
	\label{eqn:mpisq-wqandl}
	\mpisq &&= \mu_{Bare}^2 +\lambda^2\HVEV^2  \nonumber
	\\ &&
 	\quad -\left[\!\frac{-6\lambda^2\HVEV^2 \!+\! 4 \mathop{\sum}\!m_{quark}^2 \!+\! 4 \mathop{\sum}\! m_{lepton}^2}
 	{16\pi^2\HVEV^2}\! \right]\Lambda^2\nonumber 
	\\  &&\xrightarrow[\HVEV \neq 0, \mpisq\rightarrow0]{}   m^{2}_{\pi ;NGB;\LSMmth ;+SM q \& l}  
	\\ && \equiv 0 \nonumber\\
	m_h^2 &&\to 2 \lambda^2 \HVEV^2,\nonumber
	 \end{eqnarray}
	 where the sum is over all flavors, and for quarks also over all colors. Equation (\ref{eqn:mpisq-wqandl}) fixes the numerical value of $\mu_{Bare}^2$ so that there is exactly zero finite remnant after UVQD cancellation.   
	 $m_h^2$  is still the physical renormalized scalar Higgs pole mass-squared.

 {\bf 1-loop UVQD corrections to the $SU(2)_L$  PCAC relation vanish in Goldstone mode:} Following \cite{LynnStarkman2013a}, we show that the 1-loop UVQD corrections to the $SU(2)_L$ PCAC relation also vanish (without fine-tuning) in Goldstone mode:
\begin{eqnarray}
\label{PCAC1-loopUVQDSMQandLGoldstone}
\partial_{\mu} {\vec J}^{5}_{\mu} &&= -\HVEV [ \mu_{Bare}^2 +\lambda^2 \HVEV^2 \nonumber
\\ &&\quad\quad -  C^{Unrenorm;1-loop;\Lambda^2}_{\LSMmth+SM q\& l} \Lambda^2  ]~ \vec \pi \nonumber
\\  &&=- \HVEV \mpisq \vec \pi 
\\ &&\xrightarrow[\HVEV \neq 0, \mpisq\rightarrow0]{}  - \HVEV m^{2}_{\pi ;NGB;{\LSMmth+SM q\& l} }~ \vec \pi \nonumber
\\ &&\equiv 0\,. \nonumber
\end{eqnarray}
{\em But, because $SU(2)_L$ is spontaneously broken, $SU(2)_L$ chiral symmetry  is restored in the Goldstone mode  limit, so the conservation of the axial-vector current in  (\ref{PCAC1-loopUVQDSMQandLGoldstone})  cannot be considered to be  fine-tuned.}

{\bf Restoration of $SU(2)_L$ chiral symmetry in the spontaneously broken limit of the  $SU(2)_L$ \LSM~with PCAC:} Following \cite{LynnStarkman2013a}, we re-write the Goldstone mode 1-UVQD-improved effective Lagrangian in the unitary $\Phi$ representation \cite{Ramond2004,Georgi2009}:
\begin{eqnarray}
\label{eqn:VGoldstone1-loopUnitarySMQandL}
&L^{Effective;1-loop;\Lambda^2}_{Goldstone;{\LSMmth+SM q\& l;\Phi}}\!\!\!\!\!\!\!\!\!\!\!\!\!\!\!\!\!\!\!\!\!\!\!\!\!\!\!\!\!\!\!\!\!\!\!\!\!\!\!\!\!\!\!\!\!\!\!\!& \quad\quad\quad\quad\quad\quad\\
&&= -\frac{1}{2} \left( \partial_{\mu} {\tilde H} \right)^2
- \frac{1}{4} {\tilde H}^2 Tr \left[ \partial_{\mu} U^{\dagger} \partial_{\mu} U \right] \nonumber
\\ &&- \frac{1}{4} \lambda^2\left[ {\tilde H}^2 - \HVEV^2 \right]^2 \nonumber
\\ &&= L^{Effective;1-loop;\Lambda^2}_{Goldstone;{\LSMmth}}\,, \nonumber
\end{eqnarray} 
whose scalar sector has the same inhomogeneous NGB $SU(2)_L$ chiral symmetry 
${\vec{\tilde\pi}} \rightarrow {\vec{\tilde\pi}} + {\HVEV {\vec \theta}}~+ {\cal O}(\theta^2)$ as it did without the SM fermions. The ${\vec{\tilde\pi}}$ are true NGB with only derivative couplings.
{\em By definition, the restoration of an exact symmetry cannot be considered fine-tuning, so neither can the Goldstone mode results of this paper!}

{\bf Higgs mass-squared $m_h^2$ is not fine-tuned in the Goldstone mode of  the \GMLfull \LSM:} A central observation of this paper is that all finite remnants of 1-loop UVQD contributions to $m_h^2$ are absorbed into the dimension-2 relevant operator
$\frac{1}{2}m^2_{\pi;NGB;\LSMmth+SM q\& l} h^2$
in  (\ref{V1LoopLSMSMQandL}).
Therefore all 1-loop UVQD  contributions to $m_h^2$ vanish identically with $m^2_{\pi;NGB;\LSMmth} \equiv 0$. 
Re-writing (\ref{eqn:VGoldstone1-loopUnitarySMQandL}),
 \begin{eqnarray}
L^{Effective;1-loop;\Lambda^2}_{Goldstone;\LSMmth+SM q\&l; \Phi} = -\vert\partial_\mu\Phi\vert^2 &&  \\
 &&\!\!\!\!\!\!\!\!\!\!\!\!\!\!\!\!\!\!\!\!\!\!\!\!\!\!\!\!\!\!\!\!\!\!\!\!\!
 - \lambda^2\left[ \half h^2 + \half \pi_3^2 + \pi_+\pi_- + \HVEV h\right]^2  \nonumber
  \end{eqnarray}
  then gives the sensible Higgs mass
\be
\label{NoHiggs-FTPSMQandL}
  m_h^2 = 2\lambda^2 \HVEV^2 .
 \ee
$m_h^2$ is here at worst logarithmically divergent, because $\HVEV$ receives no UVQD \cite{Lee1970,Symanzik1970a,Symanzik1970b,Vassiliev1970,ItzyksonZuber1980}. But (crucially) we have traced (\ref{NoHiggs-FTPSMQandL}) to its origin, the restoration of inhomogeneous NGB $SU(2)_L$ chiral symmetry  in the Goldstone mode of the \GML \LSM. {\em Because it is the result ofrestoration of exact $SU(2)_L$ chiral symmetry, the Higgs mass-squared in (\ref{NoHiggs-FTPSMQandL}) is not fine-tuned!} The spontaneously broken limit of the $SU(2)_L$ \LSM~ (with SM quarks and leptons) therefore has no Higgs Fine Tuning Problem.

 The results in each of the other sections then generalize to include all UVQD traceable to virtual SM quarks and leptons:

{\bf Section \ref{sbscn:Higgs-FTP} (1-loop)}: 
 	At 1-loop the  generic $SU(2)_L$ \LSM~ in the   $\vert\HVEV\vert$ vs. $\mpisq/\lambda^2$  quarter-plane of Figure \ref{fig6} (i.e. away from the y-axis, $\mpisq=0$), 
	and including  Wigner mode (x-axis), has additional surviving finite remnants of UVQD due to virtual SM quarks and leptons, and continues to have a Higgs-FTP.

{\bf Section \ref{sbscn:all-loop-noHiggs-FTP} (1PI multi-loops)}:  
\begin{itemize}
\item 	Calculation of 1PI naive degrees of divergence reveals \cite{Lee1970,LynnStarkman2013a,Symanzik1970a,Symanzik1970b,Vassiliev1970,Symanzik1969,Gervais1969,Lee1969,Bogoliubov1957,Hepp1966,ZimmermanDeser}
	that UVQD can only appear in scalar 2-point and 1-point 1PI functions in the scalar-sector of the effective Lagrangian.
	It is shown elsewhere \cite{LynnStarkman2013a} that (for UVQD purposes) the Ward-Takahashi identity governing the all-loop-order 2-point pion Greens function determines the strength of symmetry breaking. That U(1) result is extended to $SU(2)_L$ in \cite{Lynn2011}: 
	\be
	\label{eqn:All-loopWTIdWithFermions}
\epsilon_{\LSMmth+SM q\& l}^{All-loop}  = \HVEV\mpisq
\ee
\item Then, following \cite{LynnStarkman2013a} but extended here to $SU(2)_L$, the Higgs Vacuum Stability Condition is automatically enforced by Ward-Takahashi identities,
so that the Higgs $\langle h \rangle =0$ everywhere in the $\vert\HVEV\vert-\mpisq/\lambda^2$ quarter-plane and the physical Higgs particle cannot simply disappear into the vacuum.
\item After imposition of the Goldstone-SRC
$m^{2}_{\pi ;NGB; \LSMmth +SM q \& l} \to 0$,
  all-loop-orders UVQD therefore vanish identically in the spontaneously broken Goldstone mode limit, including all UVQD traceable to virtual SM quarks and leptons. 
  \item The 1-loop observations enumerated at the end of Section \ref{sbscn:1loop-noHiggs-FTP} therefore generalize  to all-perturbative-loop-orders with SM quarks and leptons.

\end{itemize}

That all 1-loop UVQD due to virtual SM quarks and leptons cancel exactly in the SM Higgs self-energy and mass  after tadpole renormalization 
	has been known  \cite{Lynn1982,Stuart1985} for more than three decades.

\section{Conclusion}
 \label{secn:conclusion}

Belief in huge non-vanishing remnants of cancelled ultra-violet quadratic divergences, and consequent fine-tuning problems in field theories with fundamental scalars, 
fruitfully formed much of the original motivation for certain proposed Beyond the Standard Model (BSM) physics. A partial list would include: 
Technicolor \cite{Susskind1979}, low-energy supersymmetry \cite{DimopoulosGeorgi},  little Higgs theories \cite{LittleHiggs} and Lee-Wick theories \cite{Grinstein2007}.

Susskind \cite{Susskind1979} was probably the first to motivate BSM physics out of concern for fine-tunings in the scalar sector.   He argued that
fundamental scalars are bad because they lead to UVQD that must be fine tuned away.  The motivation for technicolor was to 
replace the fundamental scalar Higgs doublet with a composite condensate of new fermions, called techniquarks, with QCD-like confining interactions.
Fundamental scalars {\em do} result in UVQD that must be fine tuned away under many circumstances ({\it e.g.} in Section \ref{sbscn:Higgs-FTP}), and so most renormalizable field
theories with fundamental scalars do have a Higgs Fine Tuning Problem.   However, as we have shown here and elsewhere \cite{Lynn2011, LynnStarkman2013a}, the Goldstone Theorem  can protect a limited number of scalars 
(at least those in the same multiplet as the Nambu-Goldstone Bosons) from such fine-tuning problems:  we have called this effect {\em ``the Goldstone Exception"}.  We have shown that the  spontaneously broken limit of the $SU(2)_L$ Linear Sigma Model with one complex Higgs
doublet, the same UVQD structure as the Standard Model in its naive zero-gauge-coupling limit, is one example of such a theory that {\em is} protected in Goldstone mode, i.e. when spontaneously broken.
This is an important exception to the canonical objection to fundamental scalars.  More particularly, it is an exception that the stand-alone SM,
unadorned by BSM physics, makes use of \cite{Lynn2011} so that the Standard Model Higgs mass is not-fine-tuned.

Low energy supersymmetry (SUSY) aims to solve its own Higgs fine-tuning problem by the miraculous cancellation of the UVQD
portion of the loop contributions of particles against those of their superpartners  \cite{Weinberg2000,BaerTata2006,RandallReece2012,Weinberg2005,DimopoulosArkaniHamed2004}.  The key observation is that if the particle is a boson its superpartner
is a fermion, and vice versa -- the $-1$  that fermions pick up in loops relative to bosons is key  to understanding this cancellation. (But 
not sufficient -- the numerical values of the coefficients must be equal, and are as a consequence of the supersymmetry.)
Low energy SUSY might be used to solve the Higgs fine-tuning problem of theories that have one.   However, since the spontaneously broken $SU(2)_L$ \LSM~
(and the SM \cite{Lynn2011}) does not have this problem, low energy SUSY can only solve the scalar fine-tuning problems of BSM physics.
For example, certain Grand Unified Theories {\em may}  have a Higgs fine-tuning problem, as finite contributions to boson mass-squareds will be of order
the GUT scale, and may need to be fine tuned down to the weak scale.  Thus SUSYGUTs could then make use of low energy SUSY to avoid a Higgs Fine-Tuning Problem.

The role of the Goldstone Theorem in taming the Higgs fine-tuning problem in spontaneously broken O(4) \LSM~ (and also in U(1) \cite{LynnStarkman2013a}, $SU(2)_L$ with PCAC and the Standard Model \cite{Lynn2011}) 
appears to have gone unnoticed.
As we have shown, because ultra-violet quadratic divergences are sensitive only to ultraviolet physics, the O(4) symmetry forces the UVQD
contributions to the Higgs mass-squared to be identical to those of the other three bosons in its multiplet.  This is true whether  the  O(4) symmetry
is spontaneously broken, explicitly broken or unbroken.   In the unbroken (Wigner-mode) and explicitly broken phases of the theory, this
still leaves a Higgs fine-tuning problem, if the boson masses are meant to be small compared to whatever scale the UV cutoff $\Lambda$ is taken to be.  However, in the spontaneously
broken phase the Goldstone Theorem guarantees that the companions of the Higgs are exactly massless Nambu-Goldstone  Bosons, and
thus automatically eliminates any common UVQD contributions to the masses of the NGBs and the Higgs.   

We have shown, following \cite{LynnStarkman2013a} but extended here to the O(4)\LSM~(and also to $SU(2)_L$ \LSM~with SM quarks and leptons), that, including all S-Matrix UVQD, the requirement that the Higgs vacuum is automatically stable (against spontaneous appearance or disappearance of physical Higgs) is unexpectedly enforced by Ward Takahashi identities, and does not need to be imposed separately.   This is a new result, as B.W. Lee, K. Symanzik and C. Itzykson and J-B. Zuber
seem to have  believed that the Higgs Vacuum Stability Condition (vanishing of tadpoles) needed to be separately imposed.  

For the O(4) and $SU(2)_L$ \LSM s, the Higgs VEV $\HVEV$ (which receives no UVQD corrections \cite{Lee1970,Symanzik1970a,Symanzik1970b,Vassiliev1970,ItzyksonZuber1980} ) is a free parameter of the model.  In the case of the spontaneously broken Goldstone mode theory, once $\HVEV$ is chosen, the natural scale of the Higgs boson mass-squared is $\HVEV^2$, and this is stable against UVQD quantum fluctuations, 
with $m_h^2$ picking up  only finite and logarithmically divergent contributions from wavefunction renormalization.

We have shown that S-Matrix UVQD sum exactly to zero in the spontaneously broken O(4) and $SU(2)_L$  \LSM s.  
This is traced to the UVQD Goldstone-Mode  Renormalization Prescription (GMRP): 
after Ward-Takahasi-identity-enforced Higgs-VSC tadpole renormalization, 
the UVQD-corrected theory requires explicit enforcement of the Goldstone Theorem  
by imposition of Lee/Symanzik$^\prime$s Goldstone-Symmetry Restoration Condition. 
All S-Matrix UVQD, and their finite remnants, therefore vanish identically with the NGB mass: this  includes ``new" UVQD (widely unfamiliar to modern audiences) we have identified as corrections to the PCAC relation. 
Our no-fine-tuning-theorem for a weak-scale Higgs mass is therefore simply another (albeit un-familiar) consequence of the Goldstone Theorem, 
an exact property of the  O(4) and $SU(2)_L$ \LSM s vacuum and excited states. 

The vanishing of all UVQD is traced directly, in this paper, to the restoration of axial-vector current conservation and $SU(2)_{L-R}$ chiral symmetry (or $SU(2)_L$ chiral symmetry when including SM fermions). The $\vec {\tilde \pi}$ are true Nambu Goldstone bosons with only derivative couplings. {\em The restoration of an exact symmetry is considered the best definition of no-fine-tuning, so the Goldstone-mode results of this paper are not fine-tuned.} The spontaneously broken limit of the  \GMLfull \LSM~ already has sufficient symmetry to force all S-Matrix UVQD remnants to vanish identically 
and to ensure that it has  no Higgs Fine Tuning Problem.
It is un-necessary to impose any new symmetries or appeal to new ``Beyond the O(4) \LSM" physics.

The restoration of exact chiral $SU(2)_L$ symmetry in the Goldstone-mode limit of the stand-alone $SU(2)_L$ \LSM~ (i.e. one complex Higgs doublet plus SM quarks and leptons, with loop integrals cut off at some much higher UV scale $\Lambda$), protects it from that higher scale 
(up to terms proportional to $\sim\ln\Lambda$). {\em The spontaneously broken global $SU(2)_L$ \LSM~ therefore has no Higgs Fine Tuning or Naturalness problem.} 
But that may be spoiled if it is embedded/integrated into some higher scale ``Beyond the $SU(2)_L$ \LSM" theory. Still, that is not the problem of the Goldstone-mode global $SU(2)_L$ \LSM. 

The Goldstone-mode global $SU(2)_L$ \LSM~  studied above {\em is} the naive zero-gauge coupling limit of the scalar sector of the Standard Model.   
It is here that the Higgs Fine Tuning Problem in the Standard Model is generically identified.  
As shown in \cite{LynnStarkman2013a} in the context of a $U(1)$ theory,  the subtleties of gauging the global symmetry of the \LSM~ do not reintroduce a Higgs-FTP --
the UVQDs all remain subsumed in the vanishing NGB mass squared when one insists on the Goldstone Mode Renormalization Prescription of Lee, Symanzik and Taylor.
As demonstrated in \cite{Lynn2011}, for all the complications of the Standard Model (including the presence of $U(1)_Y$ and the gauging of it and the non-Abelian $SU(2)_L$) 
these results remain robust -- when one correctly imposes the Goldstone Mode Renormalization Prescription, all UVQDs are  contained in the NGB mass squared and  so vanish exactly with no finite remnants,
due to a symmetry and thus without fine tuning. The Standard Model capitalizes on the ``Goldstone Exception"  to avoid a Higgs Fine Tuning Problem.
It is up to those introducing any new high-scale physics to maintain the Goldstone Exception,  or, at worst, augment it with a naturalness mechanism of their own.

\smallskip
\smallskip
\acknowledgments
BWL thanks Jonathan Butterworth and the Departments of Physics at University College London, CERN, CWRU and Texas A\&M (where this work began) for hospitality 
and Bruce Campbell, Amir E. Mosaffa and Jennifer Thomas for valuable discussions. 
We are indebted to Raymond Stora for re-constructing the history of renormalization
 of the O(4) \LSM~  in the   $\vert\HVEV\vert$ vs. $\mpisq/\lambda^2$  quarter-plane and in the Goldstone-mode limit, and for K. Symanzik's dictum \cite{SymanzikPC}. 
 GDS and DIP are  supported by a grant from the US Department of Energy  to the theory group at CWRU.   
 KF is supported by the DOE at the University of Michigan.
 GDS and KF thank the CERN Theory group for  its hospitality.

\appendix  
\section{1-loop UVQD Lagrangian, $L^{1-loop;\Lambda^2}_{\LSMmth}$, in pure scalar O(4) \LSM, in the   $\vert\HVEV\vert-\mpisq/\lambda^2$  quarter-plane}

In this Appendix we derive the 1-loop UVQD Lagrangian of  (\ref{eqn:L-1loop-Lsq-LSM}).  We are only interested in UVQD contributions to the 1-loop effective Lagrangian: these can be evaluated at zero momentum. We obtain the two relevant operators:
the 1-loop two-point self-energy (cf. figures \ref{fig1}, \ref{fig2} and \ref{fig3}) in subsection \ref{2point}
and the 1-loop 1-point functions (cf. figure \ref{fig4}) in subsection \ref{1point}.  We follow the Feynman diagram naming conventions of Veltman 
 \cite{Veltman1981}) in defining the coefficients
$P_{9A}$ with virtual $\pi_3$,  $P_{9B}$ with virtual $\pi_\pm$,  and $P_{10}$ with virtual $h$.
 
\subsection{1-loop 2-point functions} 
\label{2point}

\begin{figure}[htpb]
\includegraphics[width=0.3\textwidth,height=0.4\textheight]{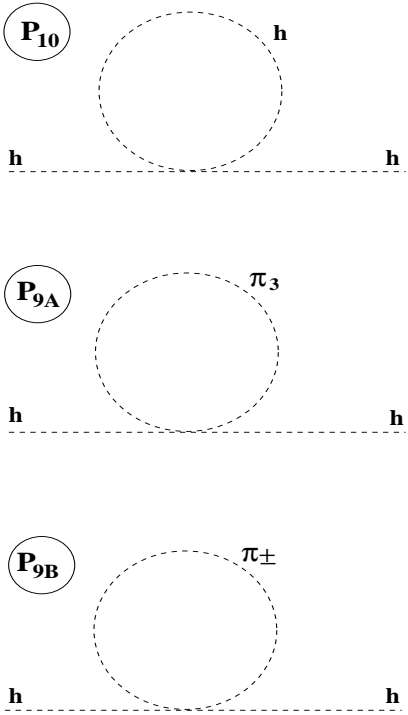}
\caption{Graphs contributing to $L^{{\rm 1-loop, 2-point};\Lambda^2}_{\LSMmth ;hh}$.}
\label{fig1}
\end{figure}

\begin{figure}[htpb]
\includegraphics[width=0.3\textwidth,height=0.4\textheight]{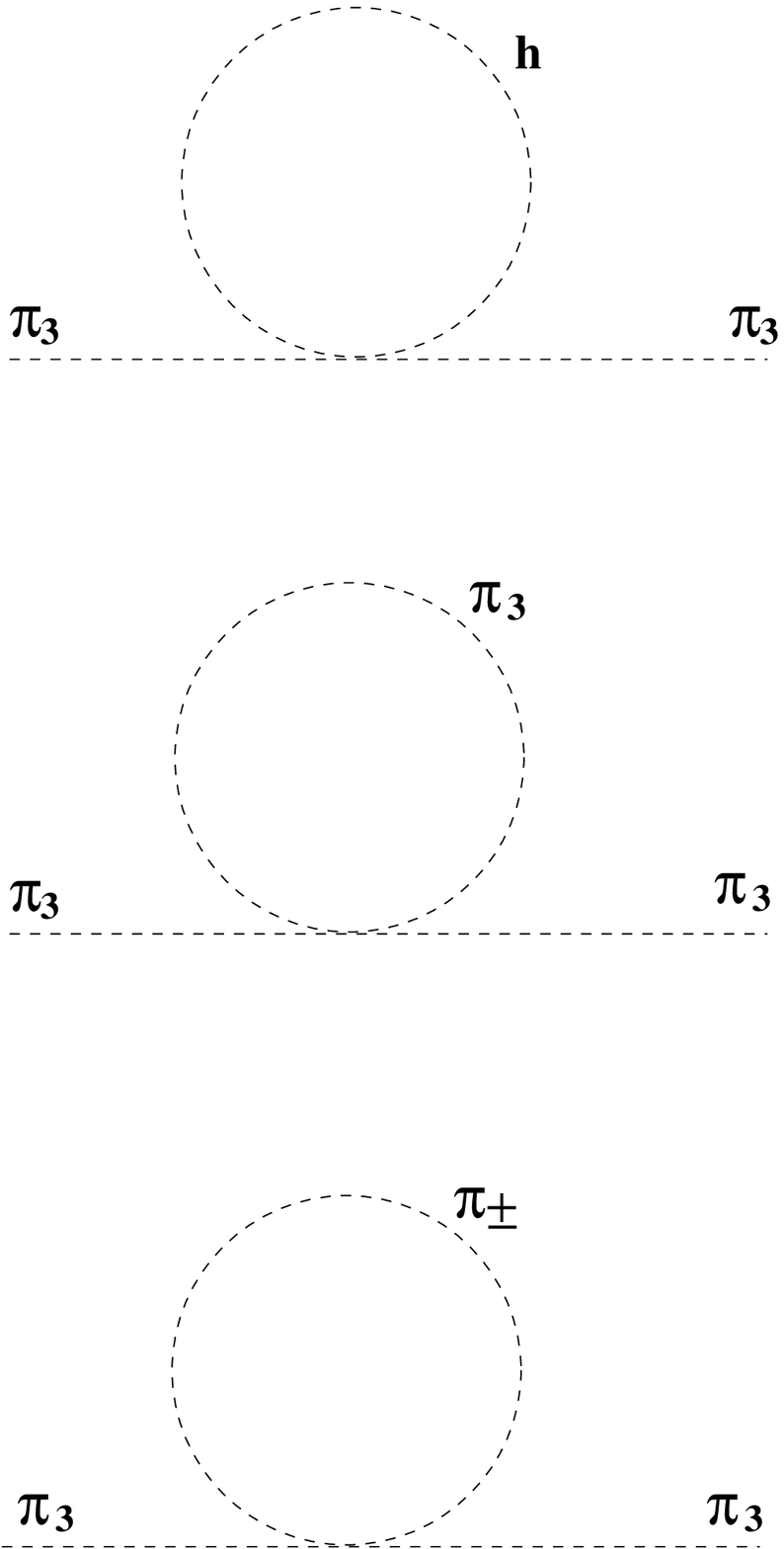}
\caption{Graphs contributing to $L^{{\rm 1-loop, 2-point};\Lambda^2}_{\LSMmth ;\pi_3\pi_3}$.}
\label{fig2}
\end{figure}

\begin{figure}[htpb]
\includegraphics[width=0.3\textwidth,height=0.4\textheight]{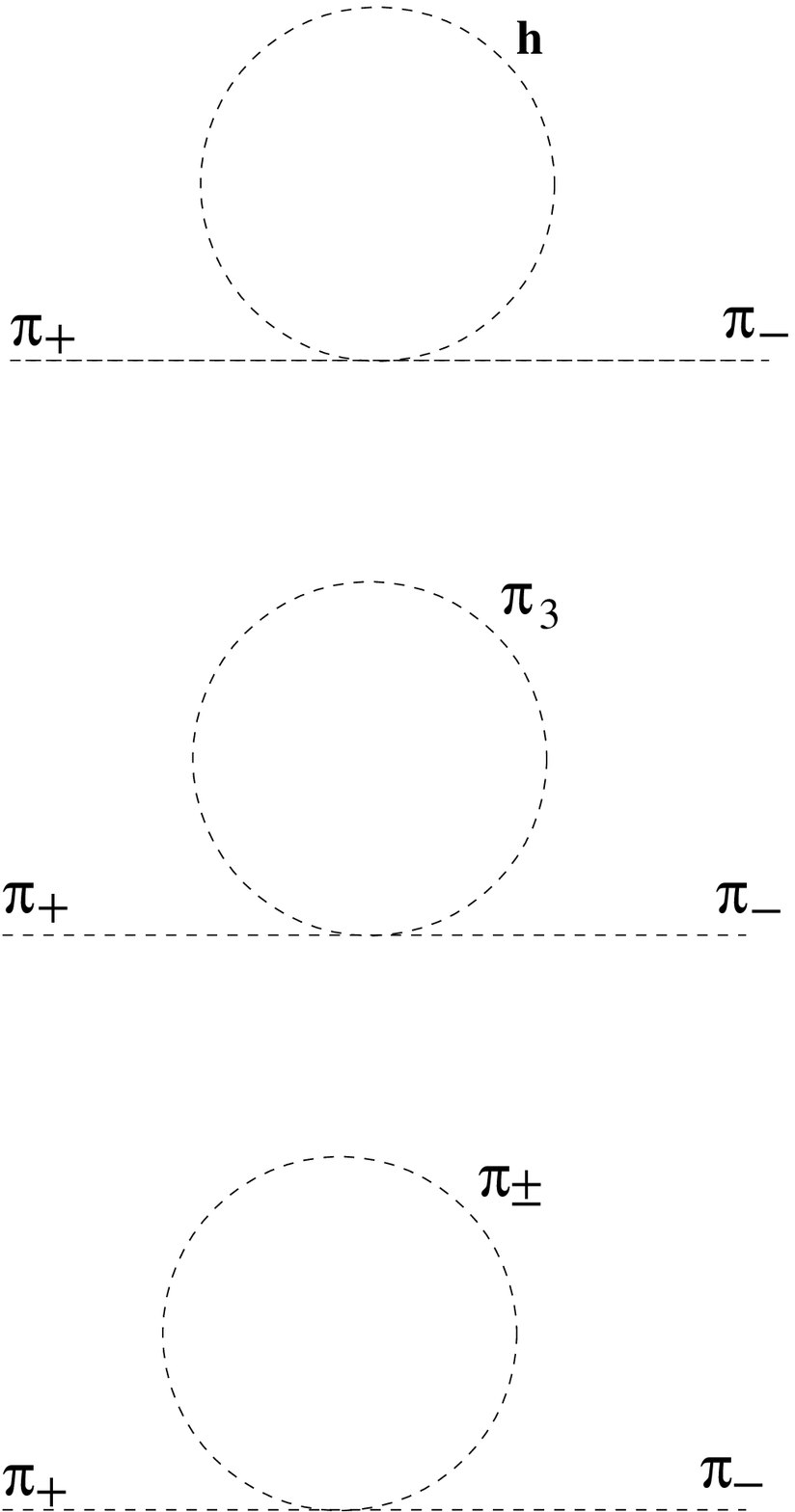}
\caption{Graphs contributing to $L^{{\rm 1-loop, 2-point};\Lambda^2}_{\LSMmth ;\pi_+\pi_-}$.}
\label{fig3}
\end{figure}

The 1-loop 2-point contributions from Higgs and pion self-energies are shown in figures \ref{fig1}, \ref{fig2} and \ref{fig3}.
In each figure, the top panel illustrates the contribution from Higgs loops; the middle panel from $\pi_3$ loops; and the bottom panel 
from $\pi_{\pm}$ loops.

The UVQD contributions of 1-loop 2-point  $hh$ Higgs self-energy diagrams \cite{Veltman1981,Lynn1982b,Stuart1985}
 in figure \ref{fig1}) to the 1-loop UVQD 2-point Lagrangian are 
\be
L^{1-loop;2-point;\Lambda^2}_{\LSMmth;hh} = \Lambda^2\left(p_{9A} + p_{9B} + p_{10}\right) \frac{h^2}{2}
\ee
with
\be
p_{9A} = -\frac{\lambda^2}{16\pi^2 }; \,\,
p_{9B} = -2\frac{\lambda^2}{16\pi^2 };  \,\,
p_{10} = -3\frac{\lambda^2}{16\pi^2 }\,.
\ee

But the UVQD contribution of {\em each} of the 1-loop 2-point  $\pi_3\pi_3$ neutral-pseudoscalar self-energy diagrams (Figure \ref{fig2})
	in $L^{1-loop;2-point;\Lambda^2}_{\LSMmth;\pi_3\pi_3}$,
and the 1-loop 2-point   charged-pseudoscalar  $\pi_-\pi_+$ self-energy diagrams  (Figure \ref{fig3}) in 
	$L^{1-loop;2-point;\Lambda^2}_{\LSMmth;\pi_-\pi_+}$
are related (by explicit calculation) to their associated 1-loop 2-point  $hh$ HiggsÕ self-energy diagrams  (Figure \ref{fig1})
	in $L^{1-loop;2-point;\Lambda^2}_{\LSMmth;hh} $
by Clebsch-Gordon coefficients and combinatorics. 

Adding all these into the UVQD 1-loop 2-point Lagrangian gives: 
\begin{eqnarray}
 L^{1-loop;2-point;\Lambda^2}_{\LSMmth;\Phi^\dagger\Phi}
   &&=   L^{1-loop;2-point;\Lambda^2}_{\LSMmth;hh}\nonumber\\ 
   &&\!\!\!\!\!\!\!\!\!\!\!\!\!\!\!\!\!\!\!\!\!\!\!\!\!\!\!\!\!\!\!\!\!\!\!\!\!
   + L^{1-loop;2-point;\Lambda^2}_{\LSMmth;\pi_3\pi_3}
   + L^{1-loop;2-point;\Lambda^2}_{\LSMmth;\pi_-\pi_+} \\
   &&\!\!\!\!\!\!\!\!\!\!\!\!\!\!\!\!\!\!\!\!\!\!\!\!\!\!\!\!\!\!\!\!\!\!\!\!\!
   =   \left(p_{9A} + p_{9B} + p_{10}\right) \Lambda^2 \left(  \frac{h^2}{2} + \frac{\pi_3^2}{2} + \pi_+\pi_-\right) \,.\nonumber
   \end{eqnarray}

\subsection{1-loop 1-point functions}
\label{1point}
\begin{figure}[htpb]
 \includegraphics[width=0.3\textwidth,height=0.4\textheight]{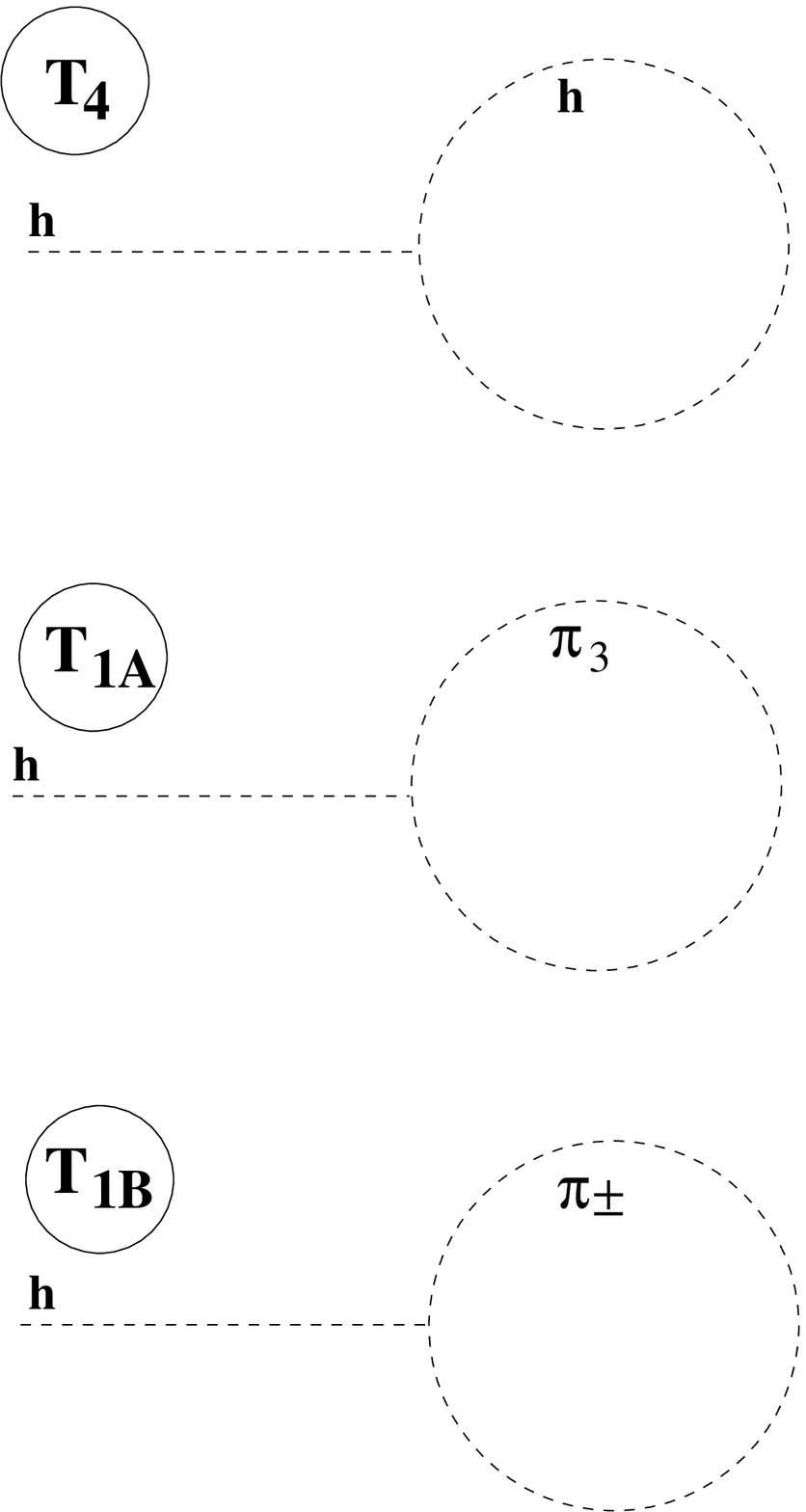}
\caption{Graphs contributing to $L^{{\rm 1-loop, 1-point};\Lambda^2}_{\LSMmth ;\langle{}H\rangle{}h}$.}
\label{fig4}
\end{figure}

 With $\HVEV^2 > 0$  the theory also receives UVQD contributions from 1-loop 1-point $\HVEV h$   tadpole diagrams  \cite{Veltman1981,Lynn1982b,Stuart1985} (Figure \ref{fig4}):  
 $T_{1A}$ with a virtual $\pi_3$ loop,   
 $T_{1B}$ with a virtual $\pi_\pm$ loop,   
 $T_{4}$ with a virtual $h$ loop).
The 1-loop tadpole diagrams (Figure \ref{fig4}) have the same structure as their 1-loop 2-point counterparts in Figure \ref{fig1}, but with one of the legs
in the 2-point diagram replaced by a Higgs VEV.

The UVQD contributions of these tadpole diagrams to S-matrix elements 
(e.g. boson propagator tadpole insertions) 
can be incorporated via a 1-loop 1-point tadpole Lagrangian
\be
L^{1-loop;1-point;\Lambda^2}_{\LSMmth;\HVEV h} =  \left( t_{1A} + t_{1B} + t_4\right) \Lambda^2 h.
\ee
But explicit calculation  \cite{Veltman1981, Lynn1982b,Stuart1985} and visual comparison of Figures \ref{fig1} and \ref{fig4}
) shows that the UVQD in {\em each} of the 1-loop 1-point tadpole diagrams 
is proportional to the UVQD in its associated 1-loop 2-point Higgs self-energy diagram:
\be
t_{1A}=p_{9A}\HVEV,\,\, 
t_{1B}=p_{9B}\HVEV,\,\, 
t_{4}=p_{10}\HVEV,
\ee
so that 
\be
L^{1-loop;1-point;\Lambda^2}_{\LSMmth;\HVEV h} = \left( p_{9A} + p_{9B} + p_{10}\right) \Lambda^2 \HVEV h.
\ee

\subsection{Combined 1-loop 1-point and 2-point functions}

Combining the 1-loop 2-point and 1-loop 1-point functions,  
we can therefore form the total UVQD 1-loop \LSMmth~ Lagrangian
\begin{eqnarray}
L^{1-loop;\Lambda^2}_{\LSMmth}&& \\
&&\!\!\!\!\!\!\!\!\!\!\!\!\!\!\!\!\!\!\!\!\!\!\!\!\!\!\!\!\!\!\!\!\!\!\!\!\!
=  L^{1-loop;2-point;\Lambda^2}_{\LSMmth;\Phi^\dagger\Phi} + L^{1-loop;1-point;\Lambda^2}_{\LSMmth;\HVEV h}\nonumber\\
&&\!\!\!\!\!\!\!\!\!\!\!\!\!\!\!\!\!\!\!\!\!\!\!\!\!\!\!\!\!\!\!\!\!\!\!\!\!
= \left( p_{9A} + p_{9B} + p_{10}\right) \Lambda^2 \left( \frac{h^2}{2} + \frac{\pi_3^2}{2} + \pi_+\pi_- + \HVEV h\right) \nonumber\\
&&\!\!\!\!\!\!\!\!\!\!\!\!\!\!\!\!\!\!\!\!\!\!\!\!\!\!\!\!\!\!\!\!\!\!\!\!\!
= C^{Unrenorm;1-loop;\Lambda^2}_{\LSMmth} \Lambda^2 \left(\Phi^\dagger\Phi - \half\HVEV^2\right)\nonumber
\end{eqnarray}
where
\begin{eqnarray}
\label{eqn:app-quote1}
 C^{Unrenorm;1-loop;\Lambda^2}_{\LSMmth} &\equiv&  \left( p_{9A} + p_{9B} + p_{10}\right)  \\
 &=& -\frac{6\lambda^2}{16\pi^2} \,.\nonumber
\end{eqnarray}
which is the result in  (\ref{eqn:L-1loop-Lsq-LSM}).

\section{UVQD vanish identically in Goldstone mode O(4) \LSM~  to all perturbative loop-orders. The Higgs mass is not fine-tuned.}

B.W.Lee and K.Symanzik renormalized pure scalar \GMLfull model to all-loop-orders throughout the  $\vert\HVEV\vert$ vs. $\mpisq/\lambda^2$  quarter-plane in the language of BPHZ
\cite{Lee1970,Symanzik1970a,Symanzik1970b,Symanzik1969,Lee1969,Bogoliubov1957,Hepp1966,ZimmermanDeser}.
Aside from vacuum bubbles (ignored here), they proved that 1PI multi-loop UVQD can only appear in the coefficients of:
\begin{itemize}
\item Scalar-sector 2-point operators proportional to $\Phi^\dagger\Phi$;
\item Scalar-sector 1-point tadpole operators, whose 1PI loops contain at least one 3-boson vertex.
\end{itemize}
Specifically, they showed that the counter-term embedded within the appropriate dimension-2 and dimension-1 relevant operators in the bare Lagrangian
\be 
L^{Bare}_{\LSMmth} \!  \sim  -\mu_{Bare}^2\!\!\left(\frac{h^2+\pi_3^2}{2}  \!+\!  \pi_+\pi_- \!+\!  \HVEV h\!\right)\!\!
\ee
are guaranteed to remove all 1PI UVQD in the S-Matrix \cite{Lee1970,Symanzik1970a,Symanzik1970b,Vassiliev1970,Symanzik1969,Gervais1969,Lee1969,Bogoliubov1957,Hepp1966,ZimmermanDeser}.
It follows that the multi-loop UVQD contribution can be written
\begin{eqnarray}
\label{eqn:LAllLoopsLambdaSquared}
L^{All-loops;\Lambda^2}_{\LSMmth} &=& \\
&&\!\!\!\!\!\!\!\!\!\!\!\!\!\!\!\!\!\!\!\!\!\!\!\!\!\!\!\!\!\!\!\!\!\!\!\!\!\!\!\!
 C^{Unrenorm;All-loops;\Lambda^2}_{\LSMmth} 
\!\!\!
\left(\frac{h^2+\pi_3^2}{2}  \!+\! \pi_+\pi_- \!+\! \HVEV h\!\right)\Lambda^2\nonumber
\end{eqnarray}
with 
$C^{\cdots}_{\LSMmth} \left(\HVEV^2,\lambda^2,\mpisq,m_h^2=\mpisq+2\lambda^2\HVEV^2\right)$ 
a finite {\em constant} (because of the dimensionality of (\ref{eqn:LAllLoopsLambdaSquared}))
that is dependent (because of nested divergences within multi-loop 1PI graphs) on the finite physical input constant parameters of the theory. 
The ratio  $\mu_{Bare}^2/\Lambda^2$ is self-consistently taken to remain finite as  $\Lambda^2\to\infty$ . 

Now form the all-loop UVQD-improved effective Lagrangian, including all-orders scalar 2-point self-energy and 1-point tadpole UVQD (but ignoring un-interesting logarithmically divergent and finite contributions and vacuum energy/bubbles):
\begin{eqnarray}
\label{eqn:yetanotherLeff}
L^{Effective;All-loop;\Lambda^2}_{\LSMmth} 
	&=& L^{Bare;\Lambda^2}_{\LSMmth}  + L^{All-loop;\Lambda^2}_{\LSMmth}  \nonumber\\
	&&\!\!\!\!\!\!\! \!\!\!\!\!\!\! \!\!\!\!\!\!\! \!\!\!\!\!\!\! 
	= -\vert\partial_\mu\Phi\vert^2 - V^{Renorm;All-loop;\Lambda^2}_{\LSMmth}
\end{eqnarray}
 where
  \begin{eqnarray}
\label{eqn:yetanotherVrenorm}
 V^{Renorm;All-loop;\Lambda^2}_{\LSMmth}&& \nonumber\\
	&&\!\!\!\!\! \!\!\!\!\! \!\!\!\!\! \!\!\!\!\!\!\!\!\!\! \!\!\!\!\!\!\!\!\!\!\!\!\!\!\! \!\!\!\!\!
	= \lambda^2\left[\frac{h^2}{2}  + \frac{\pi_3^2}{2} + \pi_+\pi_- + \HVEV h \right]^2  \\
	&& \!\!\!\!\! \!\!\!\!\! \!\!\!\!\! \!\!\!\!\! \!\!\!\!\!\!\!\!\!\! \!\!\!\!\! + m^2_{\pi}\left[\frac{h^2}{2}  + \frac{\pi_3^2}{2} + \pi_+\pi_-\right]
	+  \left(\HVEV\mpisq-\epsilon_{\LSMmth}^{All-loop}\right) h.\nonumber
\end{eqnarray}
 with $H = h+\HVEV$ and $\langle h\rangle = 0$. 
 Here
  \begin{eqnarray}
  \label{eqn:MpiAllLoop}
  m^2_{\pi} &\equiv&  \mu_{Bare}^2 - C^{Unrenorm;All-loop;\Lambda^2}_{\LSMmth} \Lambda^2
  \\ &+& \lambda^2\HVEV^2 \nonumber
  \shortintertext{and}
m_h^2 &=& \mpisq + 2\lambda^2\HVEV^2 \geq \mpisq\,.
  \end{eqnarray}
Once again, $\mpisq$, the all-loop-corrected physical renormalized pion (pole) mass-squared (and the solution to a highly non-linear equation), 
  has absorbed all UVQD, this time to all perturbative loop-orders \cite{Lee1970,Symanzik1970a,Symanzik1970b,Vassiliev1970,Symanzik1969,Gervais1969,Lee1969,Bogoliubov1957,Hepp1966,ZimmermanDeser}.

To force the theory into the spontaneously broken Goldstone mode, Lee/Symanzik$^\prime$s  {\bf the Goldstone Mode Renormalization Prescription (GMRP)}) is imposed:

{\bf Higgs Vacuum Stability Condition (tadpole renormalization):} 
\begin{itemize}
\item The exact UVQD-corrected vacuum is
defined in subsection \ref{sbscn:O4LSMquarter-plane} 
\item As remarked before, B.W. Lee proved \cite{Lee1970} that the Ward-Takahashi identity (\ref{eqn:WTidentity}) (i.e. which governs the 2-pion connnected Greens function and sets the strength of the PCAC 
relation) and the PCAC relation itself (\ref{eqn:DivergenceofAmu}) receive no UVQD corrections to all-loop-orders of perturbation theory.
Using the O(4) Ward-Takahashi identity  \cite{Lee1970,Symanzik1970a,Symanzik1970b,Vassiliev1970,ItzyksonZuber1980}: the UVQD corrections to (\ref{eqn:LPCAC}) can therefore be included in the renormalized explicit symmetry breaking term
\begin{eqnarray}
\label{eqn:SBCurrentDivergences}
\partial_\mu\vec{J}_\mu^{5} &=& -\epsilon^{All-loop}_{\LSMmth}~  {\vec \pi}
\end{eqnarray}
\be
\label{eqn:EpsilonAllLoop}
\epsilon_{\LSMmth}^{All-loop}  = \HVEV\mpisq
\ee
where, including all-orders UVQD contributions from virtual scalars (but neglecting logarithmic divergences and other finite contributions), $\mpisq$ is the physical renormalized pion (pole) mass for the three degenerate pions.
\item Following \cite{LynnStarkman2013a} but extended here to O(4), the Higgs-VSC is then automatically enforced to all UVQD loop orders, by putting (\ref{eqn:EpsilonAllLoop}) into (\ref{eqn:yetanotherVrenorm}), by Ward-Takahashi identities,
so that $\langle h \rangle =0$ everywhere in the $\vert\HVEV\vert-\mpisq/\lambda^2$ quarter-plane and the physical Higgs particle cannot simply disappear into the all-orders UVQD-corrected vacuum.
\item After Ward-Takahashi-identity-enforced Higgs-VSC tadpole renormalization, 
 the effective scalar-sector Lagrangian, 
 keeping All-loop UVQD (but ignoring un-interesting logarithmic divergences, finite parts and vacuum energy/bubbles) 
 can again be written, in analogy with subsection \ref{sbscn:1loopUVQD},
\begin{eqnarray}
\label{eqn:VAllLoop}
V^{Renorm;All-loops;\Lambda^2}_{\LSMmth} &=& \lambda^2 \left[ \Phi^\dagger\Phi - \half\left( \HVEV^2 - \frac{\mpisq}{\lambda^2}\right)\right]^2\nonumber \\
&&- \HVEV \mpisq h
\end{eqnarray}

\end{itemize}

{\bf Goldstone Symmetry Restoration Condition:} is discussed in subsection \ref{sbscn:all-loop-noHiggs-FTP}.

\section{Effective $SU(2)_L$ Lagrangian  in the   $\vert\HVEV\vert-\mpisq/\lambda^2$  quarter-plane. 1-loop UVQD contributions from virtual SM quarks and leptons.} 

UVQD contributions to the 1-loop effective Lagrangian can be calculated at zero momentum.
Including virtual SM quarks and leptons (Reference \cite{Veltman1981}: graphs $P_{11;i}$ where $i$ runs over these fermions), 1-loop 2-point function $hh$  HiggsÕ self-energy diagrams (shown for the representative third-generation quarks and leptons in Figure \ref{fig7}):  
\begin{figure}[htpb]
\includegraphics[width=0.3\textwidth,height=0.4\textheight]{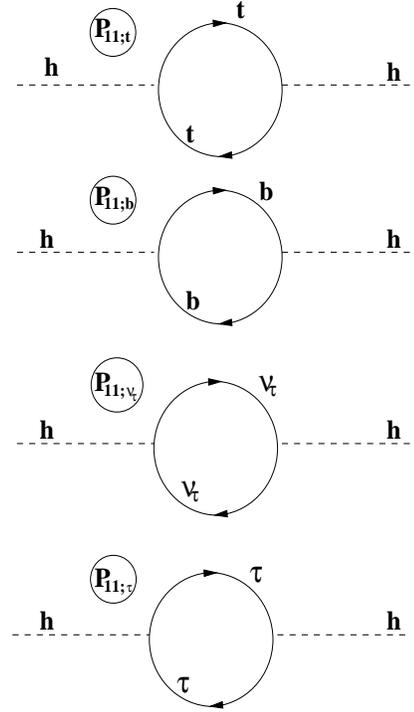}
\caption{Graphs of third-generation quarks and leptons  contributing to $L^{{\rm 1-loop, 2-point};\Lambda^2}_{\LSMmth +{\rm SM q \& l};hh}$.
There are identical graphs for the other two generations.}
\label{fig7}
\end{figure}
\be
L^{1-loop;2-point;\Lambda^2}_{\LSMmth+SM q\& l;hh} = \Lambda^2\left( p_{9A} + p_{9B} + p_{10} + p_{11}\right)\half h^2
\ee
with
\be
p_{11} = \mathop{\sum}_{i=fermions} p_{11;i} .
\ee
Here 
\be
8\pi^2 p_{11;i} = y_i^2
\ee
is the square of the Yukawa coupling to the ith fermion \cite{Veltman1981,Lynn1982b,Stuart1985}.

\begin{figure}[htpb]
\includegraphics[width=0.3\textwidth,height=0.4\textheight]{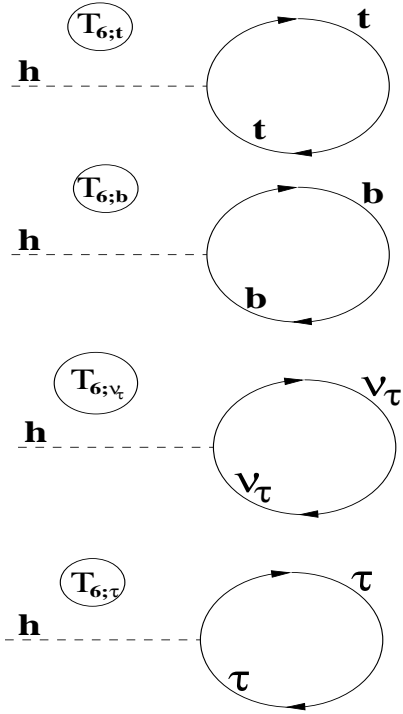}
\caption{Graphs of third-generation quarks and leptons  contributing to $L^{{\rm 1-loop, 1-point};\Lambda^2}_{\LSMmth +{\rm SM q \& l};\langle{}H\rangle{}h}$.
There are identical graphs for the other two generations.}
\label{fig8}
\end{figure}

With $\HVEV^2>0$, the theory also receives 1-loop 1-point function $\HVEV h$   tadpole diagrams  \cite{Veltman1981,Lynn1982b,Stuart1985}
(Ref. \cite{Veltman1981} naming conventions:  graphs $T_{6;i}$, where $i$ runs over all virtual SM quarks and leptons). 
These are shown for the  representative third-generation quarks and leptons in Figure \ref{fig8}.
Their contribution to S-matrix elements (e.g. boson and fermion propagator tadpole insertions) can be incorporated via a 1-loop 1-point tadpole Lagrangian
\be
L^{1-loop;1-point;\Lambda^2}_{\LSMmth+SM q\& l;\HVEV h} = \left( t_{1A} + t_{1B} + t_{4} + t_{6}\right) \Lambda^2 h\,.
\ee
But explicit calculation shows  \cite{Veltman1981,Lynn1982b,Stuart1985}
that the UVQD in each of the 1-loop 1-point tadpole diagrams in Figure \ref{fig8} 
is proportional to the UVQD in its associated 1-loop 2-point Higgs self-energy diagram in Figure \ref{fig7}, so that
\be
L^{1-loop;1-point;\Lambda^2}_{\LSMmth+SM q\& l;\HVEV h} =\left( p_{9A} + p_{9B} + p_{10} + p_{11} \right)  \Lambda^2 \HVEV h\,.
\ee

\begin{figure}[htpb]
\includegraphics[width=0.3\textwidth,height=0.4\textheight]{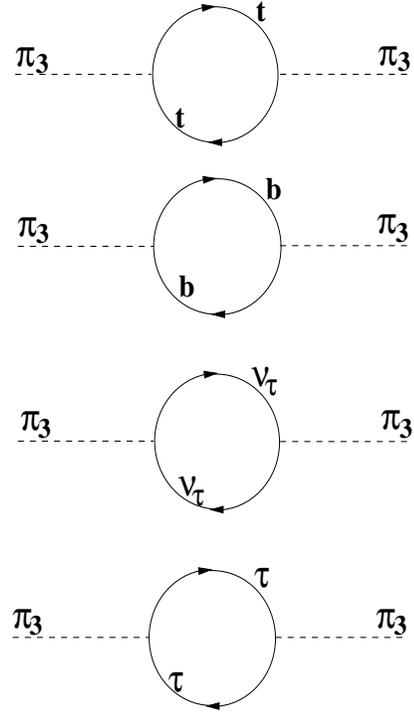}
\caption{Graphs of third-generation quarks and leptons  contributing to $L^{{\rm 1-loop, 2-point};\Lambda^2}_{\LSMmth +{\rm SM q \& l};\pi_3\pi_3}$.
There are identical graphs for the other two generations.}
\label{fig9}
\end{figure}

\begin{figure}[htpb]
\includegraphics[width=0.4\textwidth,height=0.25\textheight]{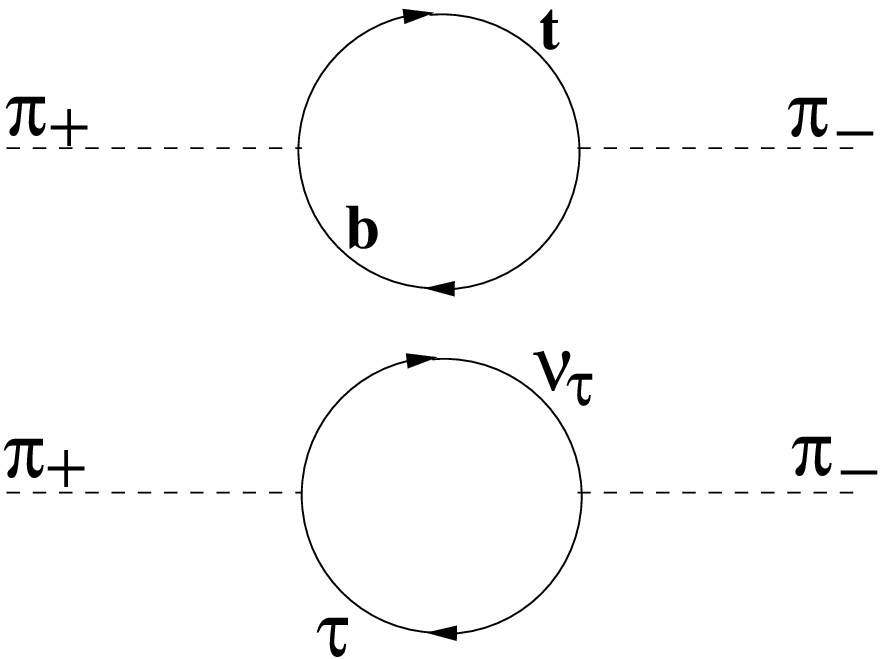}
\caption{Graphs of third-generation quarks and leptons  contributing to $L^{{\rm 1-loop, 2-point};\Lambda^2}_{\LSMmth +{\rm SM q \& l};\pi_+\pi_-}$.  
There are identical graphs for the other two generations.}
\label{fig10}
\end{figure}

Furthermore,  the UVQD contribution of each of the 1-loop 2-point  $\pi_3\pi_3$ neutral pseudoscalar self-energy diagrams (Figure \ref{fig9})
	in $L^{1-loop;2-point;\Lambda^2}_{\LSMmth+SM q\& l;\pi_3\pi_3}$, 
and of each of  the 1-loop 2-point $\pi_+\pi_-$  charged pseudoscalar self-energy diagrams (Figure \ref{fig10}) 
	in $L^{1-loop;2-point;\Lambda^2}_{\LSMmth+SM q\& l;\pi_+\pi_-}$, 
are related (by explicit calculation) to their associated 1-loop 2-point  $hh$ HiggsÕ self-energy diagrams (Figure \ref{fig7})
	in $L^{1-loop;2-point;\Lambda^2}_{\LSMmth+SM q\& l;hh}$
	by Clebsch-Gordon coefficients and combinatorics. The total UVQD 1-loop  Lagrangian is
\begin{eqnarray}	
\label{eqn:fermioncontribution}
L^{1-loop;\Lambda^2}_{\LSMmth+SM q\& l} &=& \left( p_{9A} + p_{9B} + p_{10} + p_{11} \right) \\
&&   \!\!\!\!\!\!\!\!\!\!\!\!\!\!\!\!\!\!\!\!\!\!\!\!\!\!\!\!\!\!\!\!\!\!\!
\times \Lambda^2 \left(\half h^2 + \half \pi_3^2 + \pi_+\pi_- + \HVEV h\right)\nonumber\\
&&\!\!\!\!\!\!\!\!\!\!\!\!\!\!\!\!\!\!\!\!\!\!\!\!\!\!\!\!\!\!
= C^{Unrenorm;1-loop;\Lambda^2}_{\LSMmth+SM q\& l}  \Lambda^2 \left(\Phi^\dagger\Phi - \half\HVEV^2\right)\nonumber
\shortintertext{with}
C^{Unrenorm;1-loop;\Lambda^2}_{\LSMmth+SM q\& l}  &=& \left( p_{9A} + p_{9B} + p_{10} + p_{11} \right)\,. \nonumber
\end{eqnarray}
 a    constant. 
With quark and lepton masses-squared $m_i^2 = \half y_i^2 \HVEV^2$, the coefficient in (\ref {eqn:fermioncontribution})
 can famously \cite{Veltman1981} be written
\begin{eqnarray}
\label{eqn:app-to-quote-2}
 C^{Unrenorm;1-loop;\Lambda^2}_{\LSMmth+SM q\& l}  \Lambda^2&\\
 & \!\!\!\!\!\!\!\!\!\!\!\!\!\!\!\!\!\!\!\!\!\!\!\!\!\!\!\!\!\!\!\!\!\!\!
 = \left(  \frac{  -3 m_h^2 + 4\mathop{\sum}_{quarks}m_q^2  + 4\mathop{\sum}_{leptons}m_l^2 }{16\pi^2\HVEV^2}\right)\Lambda^2\nonumber
\end{eqnarray}
where the sums are over flavor and, for the quarks, color.

Now form the scalar sector 1-loop-UVQD-improved effective  Lagrangian,
 including 1-loop 2-point self-energies and 1-loop 1-point tadpole UVQD 
 (but ignoring un-interesting logarithmically divergent and finite contributions and vacuum energy/bubbles):
 \begin{eqnarray}
 \label{LEffectiveSMFermions}
L^{Effective;\Lambda^2}_{\LSMmth+SM q\&l; \Phi} &&  \\
 &&\!\!\!\!\!\!\!\!\!\!\!\!\!\!\!\!\!\!\!\!\!\!\!\!\!\!\!\!\!\!\!
 =  L^{Bare;\Lambda^2}_{\LSMmth+SM q\&l; \Phi}  + 
 L^{1-loop;\Lambda^2}_{\LSMmth+SM q\&l}  \nonumber \\ 
 &&\!\!\!\!\!\!\!\!\!\!\!\!\!\!\!\!\!\!\!\!\!\!\!\!\!\!\!\!\!\!\!
 = -\vert\partial_\mu\Phi\vert^2 - 
	V^{Renorm;1-loop;\Lambda^2}_{\LSMmth+SM q\&l}\nonumber
\end{eqnarray}
with $H = h + \HVEV; \, \langle h \rangle = 0 $ and
\begin{eqnarray}
 \label{VEffectiveSMFermions}
V^{Renorm;1-loop;\Lambda^2}_{\LSMmth+SM q\&l} &=&
\lambda^2\left[ \half h^2 + \half \pi_3^2 + \pi_+\pi_- + \HVEV h\right]^2  \nonumber
\\ &+& \mpisq \left[ \half h^2 \!+\! \half \pi_3^2 \!+\! \pi_+\pi_- \right]  \nonumber
\\ &+&  \left(\HVEV\mpisq-\epsilon_{\LSMmth + SM q\&l}^{1-loop}\right)~ h
 \end{eqnarray}
 
 Following \cite{Lynn2011}, we set the strength of symmetry breaking by writing the $SU(2)_L$ Ward-Takahashi identity governing the pion 2-point function:
 \be
 \label{WTAllLoopSMFermions}
 \epsilon_{\LSMmth + SM q\&l}^{1-loop}=\HVEV\mpisq
 \ee
 where $\mpisq$ is the physical renormalized pion (pole) mass. 
 
 Extending the work of  \cite{LynnStarkman2013a}  to $SU(2)_L$, and including all 1-loop UVQD from virtual scalars as well as SM fermions, the Ward-Takahashi identity (\ref{WTAllLoopSMFermions}) automatically enforces the Higgs-VSC in (\ref{VEffectiveSMFermions}):
 
\begin{eqnarray}
\label{V1LoopLSMSMQandL}
 V^{Renorm;1-loop;\Lambda^2}_{\LSMmth+SM q\&l} &=&
\lambda^2\left[ \half h^2 + \half \pi_3^2 + \pi_+\pi_- + \HVEV h\right]^2  \nonumber
\\ &+& \mpisq \left[ \half h^2 \!+\! \half \pi_3^2 \!+\! \pi_+\pi_- \right] 
\end{eqnarray}

\section{1-loop n-dimensional regularization mapped onto UV cut-off regularization. Our 1-loop Goldstone mode O(4) and $SU(2)_L$ \LSM~ results are independent of regularization scheme \cite{Lynn2011}}

M.J.G.Veltman proved more than 30 years ago \cite{Veltman1981} 
 that the Passarino and Veltman $A(m^2)$ and $B_{22}(q^2,m_1^2,m_2^2)$  functions \cite{Passarino1979,Consoli1979} 
 can be made to properly capture and represent 1-loop UVQD in the SM 
 (if certain subtleties are obeyed) 
 in  n-dimensional regularization (i.e. with poles at $n=2$)
\begin{eqnarray}
 A(m^2) &\equiv& \int \frac{d^nk}{i\pi^2} \frac{1}{k^2+m^2}\\
 B_{\mu\nu}; B_{\mu}; B_{0} &\equiv& \int \frac{d^nk}{i\pi^2} \frac{k_\mu k_\nu; k_{\mu}; 1}{(k^2+m_1^2)((k+q)^2+m_2^2)}\nonumber
 \\ B_{\mu\nu} &=& \delta_{\mu\nu} B_{22} + q_\mu q_\nu  B_{21} \nonumber
  \\ B_{\mu} &=&  q_\mu  B_{1} \nonumber   
  \\ B_{3} &=&  B_{21} +B_{1} 
\end{eqnarray}
 
 On the other hand, these 1-loop integrals can be mapped to an ultra-violet cut-off $\Lambda$:       
 \be
 \frac{2}{4-n} -\gamma-\ln{\pi} \xrightarrow{\,\,n\to4\,\,} \ln{\Lambda^2}
\ee
so that 1-loop UVQD are defined
 \begin{eqnarray}
 \label{AIntegral}
 && A(m^2) \xrightarrow{\,\,n\to4\,\,}  \Lambda^2 - m^2 \left(\ln\left( \frac{\Lambda^2}{m^2}\right)+1\right)
\end{eqnarray}
(and $B_{22}$ is also defined) to self-consistently match the logarithmic  divergences and finite parts of
\begin{eqnarray}
\label{BIntegrals}
&& B_0 \left( q^2 , m_{1}^2 ,m_{2}^2 \right) \xrightarrow{\,\,n\to4\,\,}\ln{\Lambda^2} \nonumber
\\ && - \int^1_0 dx\ln{\left[ m_{1}^2 +\left( q^2 +m_{1}^2 -m_{1}^2 \right)x-x^2 q^2 -i\epsilon \right]}
\\ && B_1 \left( q^2 , m_{1}^2 ,m_{2}^2 \right) \xrightarrow{\,\,n\to4\,\,}-\frac{1}{2}\ln{\Lambda^2} \nonumber
\\ && + \int^1_0 xdx\ln{\left[ m_{1}^2 +\left( q^2 +m_{1}^2 -m_{1}^2 \right)x-x^2 q^2 -i\epsilon \right]} \nonumber
\\ && B_{21} \left( q^2 , m_{1}^2 ,m_{2}^2 \right) \xrightarrow{\,\,n\to4\,\,} \frac{1}{3}\ln{\Lambda^2} \nonumber
\\ && - \int^1_0 x^2 dx\ln{\left[ m_{1}^2 +\left( q^2 +m_{1}^2 -m_{1}^2 \right)x-x^2 q^2 -i\epsilon \right]} \nonumber
 \end{eqnarray}

The weak demand is then made, i.e. of any well-defined UV regularization scheme, that one can self-consistently change variables of integration. 1-loop UVQD then cancel exactly (with zero finite remnant) in expressions where they do not truly arise: e.g.
\begin{eqnarray}
\label{DeltaMuNuBMuNuMinusA}
 && \delta_{\mu\nu}B_{\mu\nu}+m_1^2 B_0 -A \left( m^2_2 \right) \nonumber
 \\ &&\equiv  \int \frac{d^nk}{i\pi^2} \left[ \frac{ \delta_{\mu\nu}k_\mu k_\nu +m^2_1 }{(k^2+m_1^2)((k+q)^2+m_2^2)} -\frac{1}{(k^2+m_2^2)} \right] \nonumber
 \\ && = 0
\end{eqnarray}
The finite parts of $A(m^2 )$  in (\ref{AIntegral}) and  $B_{22}$  are fixed so that changing integration variables is consistent with the expression (\ref{DeltaMuNuBMuNuMinusA}), and more generally that expressions where UVQD do not arise, such as
\be
A(m^2_1 )-A(m^2_2 ) \equiv (m_2^2 -m_1^2) B_0 (0, m_1^2 ,m_2^2 )
\ee
are true. 

Because all 1-loop UVQD vanish identically in the spontaneously broken O(4) or $SU(2)_L$ \LSM, the 1-loop results here, and  in \cite{LynnStarkman2013a,Lynn2011}, do not depend on whether n-dimensional regularization or UV cut-off regularization is used. It may be possible to push the regularization-scheme independence of our UVQD results to multi-loop 1PI operators.

 \end{document}